\documentclass[aps,nofootinbib]{revtex4-1}
\usepackage[latin1]{inputenc}
\usepackage{aas_macros}
\usepackage{amsmath}
\usepackage{amsfonts}
\usepackage{amssymb}
\usepackage{graphicx}
\usepackage{hyperref}
\usepackage{natbib}
\usepackage{xcolor}

\newcommand{\sub}[1]{_{\textrm{\tiny #1}}}

\begin{document}
\author{Mihael Peta\v{c}}
\email{mpetac@sissa.it}
\author{Piero Ullio}
\email{ullio@sissa.it}
\affiliation{SISSA, Via Bonomea, 265, 34136 Trieste, Italy}

\title{Two-integral distribution functions in axisymmetric galaxies: implications for dark matter searches}

\begin{abstract}
	We address the problem of reconstructing the phase-space distribution function for an extended collisionless system, with known density profile and in equilibrium within an axisymmetric gravitational potential. Assuming that it depends on only two integrals of motion, namely the energy and the component of the angular momentum along the axis of symmetry $L_z$, there is a one-to-one correspondence between the density profile and the component of the distribution function that is even in $L_z$, as well as between the weighted azimuthal velocity profile and the odd component. This inversion procedure was originally proposed by Lynden-Bell and later refined in its numerical implementation by Hunter \& Qian; after overcoming a technical difficulty, we apply it here for the first time in presence of a strongly flattened component, as a novel approach of extracting the phase-space distribution function for dark matter particles in the halo of spiral galaxies. We compare results obtained for realistic axisymmetric models to those in the spherical symmetric limit as assumed in previous analyses, showing the rather severe shortcomings in the latter. We then apply the scheme to the Milky Way and discuss the implications for the direct dark matter searches. In particular, we reinterpret the null results of the Xenon1T experiment for spin-(in)dependent interactions and make predictions for the annual modulation of the signal for a set of axisymmetric models, including a self-consistently defined co-rotating halo.
\end{abstract}

\keywords{two-integral phase-space distribution --- spiral galaxies --- direct detection}

\maketitle
	
\tableofcontents

\section{Introduction}

In recent years considerable efforts have been dedicated to the problem of identifying the nature of the dark matter component of the Universe. While a clean evidence for a signal is still missing, direct and indirect detection methods have set stringent constraints on several particle dark matter scenarios. Along with the improvement in experimental sensitivities, it is becoming increasingly urgent to refine theoretical predictions and properly assess various systematic errors to allow for an unbiased comparison of results from complementary techniques. Severe limitations to accurate theoretical predictions often stem from the difficulty of projecting the available observational and theoretical insights regarding dark matter halos into the underlying particle distribution models. Most critical are the cases in which the dark matter signals depend on the full phase-space distribution structure, such as when estimating the scattering probability of galactic dark matter on a target nucleus in a detector on the Earth (for a review, see, e.g.,  \cite{Green:2017odb}) or when considering annihilation probabilities depending on the relative velocity in the particle pair (see, e.g., \cite{Ferrer:2013cla,Boddy:2017vpe,Lu:2017jrh,Bergstrom:2017ptx,Petac:2018gue,Lacroix:2018qqh,Boddy:2018ike}).

The impact on direct detection has been under the closest scrutiny and it is timely to investigate it further today, given that the information on local dynamical tracers in the Galaxy is getting much richer with the data that the Gaia satellite is collecting~\cite{2018A&A...616A...1G}. While the vast majority of theoretical estimates and experimental analyses rely on approximating dark matter particle velocity distribution as a truncated isotropic Maxwelian, inspired by the configuration valid for an isolated isothermal sphere~\cite{1939isss.book.....C} and sometimes referred to as standard halo model (SHM), it has been long recognized that this is a idealized model that fails to account for important effects, such as: deviations from isotropy~\cite{Ullio:2000bf}, deviations from spherical symmetry~\cite{Evans:2000gr} and non-thermalized components with peculiar velocity patterns~\cite{Freese:2003na} (for a phenomenological update on the SHM incorporating recent inputs from Gaia, see ~\cite{Evans:2018bqy}). To the extremes in possible attempts to go beyond the SHM, one can refer to numerical realizations of Milky Way-like galaxies, see, e.g., recently~\cite{Bozorgnia:2016ogo,Kelso:2016qqj,Sloane:2016kyi,Bozorgnia:2017brl}, or directly mutuate dark matter velocity distributions from observed properties of halo stars \cite{Herzog-Arbeitman:2017fte,Herzog-Arbeitman:2017zbm,Necib:2018iwb}.  There are shortcomings in both approaches: While in a given simulation it is not straightforward to match local dynamical observables and reach the required resolution to properly account for all relevant effects, the assumptions in the second approach that the dark matter population is mostly non-thermal and properly traced by a given class of metal-poor stars are rather strong and at odd with the fact that density profiles for dark matter and halo stars appears to be significantly different, see also the discussion in~\cite{Evans:2018bqy}.

We follow here a third route, based instead on the hypothesis that the local dark matter population has reached equilibrium within the underlying gravitational potential well, and hence retrieving its distribution function from the collisionless Boltzmann equation~\cite{BinneyTremaine2008}. This route becomes particularly convenient if one introduces the simplifying assumption that all components in the system are spherically symmetric: In this case the ergodic dark matter distribution function depends only on energy, is unique and can be numerically computed via the so-called Eddington's inversion formula~\cite{BinneyTremaine2008}. The latter requires as inputs only the dark matter density profile and the overall gravitational potential profile, which can be both opportunely tuned to available dynamical tracers and other observables. This method, as well as its generalization to some classes of anisotropic distribution functions, has been applied -- within the spherical symmetry approximation -- to the Milky Way and used for direct and indirect dark matter detection studies, see, e.g.,~\cite{Ullio:2000bf,Vergados:2002hc,Catena:2011kv,Pato:2012fw,Catena2013,Ferrer:2013cla,Lacroix:2018qqh}. 

The method has several virtues: Distribution function and dark matter density profile are a priori self-consistent; The velocity profile can be directly matched onto Milky Way observables, with no need for normalizing to some estimated value of the local velocity dispersion or imposing by hand a sharp cutoff to a given local escape velocity (both being an output of the model); Within this approach, scans of the parameter space connected to, e.g., the Galactic mass model decomposition, on one hand, are much less onerous than through numerical realizations, and, on the other, produce more realistic uncertainty estimates than with phenomenological models, having an implicit embedding of the cross-correlation among relevant quantities (compare, e.g., the impact on direct detection estimated in~\cite{Catena:2011kv} to the one in, e.g.,~\cite{Green:2010gw}; despite considering analogous sets of dynamical constraints, in the latter the uncertainties on the local halo density, the local dark matter velocity dispersion and the local escape velocity need to be treated as independent quantities). The main drawback is that this method cannot account for eventual dark matter components that have not thermalized and might play a relevant role for phenomenology, see, e.g., most recently~\cite{OHare:2018trr}.

A second major issue -- which we wish to address in this paper -- is the fact that applying an inversion method devised for spherical systems to the Milky Way, and in particular to the local neighborhood or at even smaller galactocentric distances, is more motivated at the level of coping with a technical difficulty rather than on physical grounds. The local potential well is largely dominated by the stellar disc, inducing a vertical gradient much larger than the radial gradient. This affects all components of the system in dynamical equilibrium, including the dark matter halo, generating a pressure that particles feel in the azimuthal direction which is different from the pressure in the meridional plane. We therefore dismiss the approximation of ``spherical disc", and consider instead an axisymmetric environment embedding the dark halo (that in turn can be oblate/prolate rather than spherical): If one restricts to distribution functions depending only on energy and the component of the angular momentum parallel to the axis of symmetry (rather than also to the third integral of motion, rarely known), there is still a one-to-one correspondence between density profile and distribution function for a non-rotating collisionless population in equilibrium within a given axisymmetric gravitational potential. The extension of Eddington's formula to axisymmetric systems was worked out by Lynden-Bell~\cite{1962MNRAS.123..447L} and involves inverting Laplace transforms of the density with analytic continuations in the complex plane. It has been used to find few analytic distribution functions from some rather specific density profiles~\cite{hunter1975,lake1981,Dejonghe1986,Dejonghe1988,Evans1990}. Later, Hunter \& Qian~\cite{HQ1993} improved it to make it more tractable from a numerical point of view, but even in this second version there have been very few applications~\cite{HQ1995}, in particular none involving highly flattened components, possibly in connection to a technical issue encountered and solved while developing this project.

We present here for the first time two-integral-of-motion distribution functions describing an extended axisymmetric component in equilibrium within a gravitational potential getting a major contribution from a thin axisymmetric disc. This applies to the dark matter halo of a spiral galaxy, as discussed in Section~\ref{sec:axi_modelling}, but can also be readily extended to stellar halo populations. In Section~\ref{sec:MW_model} we focus on the Milky Way dark matter halo, tuning the model to dynamical observables. In Section~\ref{sec:DD}, as a first application of the developed formalism, we illustrate the impact on the predictions for dark matter direct detection rates. Finally, Section~\ref{sec:conclusions} contains our conclusions.  
	
\section{Self-consistent axisymmetric modelling}
\label{sec:axi_modelling}

The latest astronomical surveys are providing a wealth of new data, making it possible to study the dynamics within galaxies with unprecedented precision. By combining various dynamical observables, it is possible to reconstruct gravitational potentials of galaxies with ever-increasing accuracy. While photometric and radio observations allow to determine in fine details the morphology of the stellar and gas components, mass model decompositions are getting more constrained and enable to improve the estimates of the distribution of dark matter within the studied objects. In turn, these precise determinations of gravitational potential $\Phi$ and dark matter density profile $\rho$ can be exploited to reconstruct the full phase-space distribution of relaxed collisionless galactic components by means of Boltzmann equation.

In the spherically symmetric limit, the phase-space distribution function (PSDF) for isotropic systems depends only on one integral of motion -- the energy -- and can be retrieved from the density profile making use of the well known Eddington's inversion formula (see e.g.~\cite{BinneyTremaine2008}):
\begin{align}\label{eqn:eddington}
	f\sub{Edd}(\mathcal{E}) = \frac{1}{\sqrt{8} \pi^2} \cdot \frac{\mathrm{d}}{\mathrm{d} \mathcal{E}} \int_0^\mathcal{E} \frac{\mathrm{d} \Psi}{\sqrt{\mathcal{E} - \Psi}} \cdot \frac{\mathrm{d} \rho}{\mathrm{d} \Psi} \; .
\end{align}
In this formula, the relative energy $\mathcal{E} \equiv \Psi(r) - \frac{v^2}{2}$ has been introduced in terms of the relative potential $\Psi(r) \equiv \Phi_{b}-\Phi(r)$, having chosen the boundary term $\Phi_{b}$ in such way that $f\sub{Edd}>0$ for $\mathcal{E} >0$ and $f\sub{Edd}=0$ for $\mathcal{E} \le 0$. There are a few generalizations extending Eddington's inversion to anisotropic systems, such as the Ospikov-Merritt or constant-$\beta$ models~\cite{BinneyTremaine2008}, however their applicability is rather limited. Furthermore, the assumption of spherical symmetry seems oversimplifying, particularly when addressing rotationally supported galaxies, that are characterized by their stellar disc. Our work is dedicated to applying a generalization of Eddington's approach to axisymmetric systems, with the aim of studying the DM velocity distribution within spiral galaxies. It allows us to find a stationary solution of the collisionless Boltzmann equation for an arbitrary axisymmetric gravitational potential and density distribution. This is particularly interesting in the light of direct DM searches, since the local galactic potential is strongly influenced by the stellar disc, but can be in principle used to study other collisionless component of spiral galaxies.

\subsection{Phase-space inversion for axisymmetric systems}

According to the strong formulation of Jeans theorem, for a system with regular non-resonant orbits, any steady-state solution of the collisionless Boltzmann equation in a given stationary gravitational potential depends on up to three independent integrals of motion. For an axisymmetric configuration, the isolating integrals are the energy, the component of the angular momentum parallel to the axis of symmetry, $L_z$, and a so-called non-classical third integral $I_3$, which however takes an analytic expression only in very few specific cases. Hence, most often, PSDFs for axisymmetric systems have been assumed to depend on the first two only; while this is a limitation in our analysis, still it is sufficient to address the shortcomings of the spherical symmetry approximation and for building much more realistic models. 

Under the two-integral-of-motion assumption, the PSDF can be decomposed in two parts, $f_+$ that is even in $L_z$ and the $f_-$ that is odd:
\begin{align}
f(\mathcal{E}, L_z) = f_+(\mathcal{E}, L_z) + f_-(\mathcal{E}, L_z) \; .
\end{align}
The even part contains information regarding the density distribution, while the odd part describes the rotational properties of the considered system, given that:
\begin{align} 
	\rho(R, z) \equiv & \int_{|\vec{v}| \leq \sqrt{2 \Psi(R, z)}} \mathrm{d}^3 v \; f(\mathcal{E}, L_z) = \frac{2 \pi}{R} \int_{0}^{\Psi(R, z)} \mathrm{d} \mathcal{E} \int_{-R \sqrt{2(\Psi(R, z) - \mathcal{E})}}^{R \sqrt{2(\Psi(R, z) - \mathcal{E})}} \mathrm{d}L_z \; f_+(\mathcal{E}, L_z) \label {eqn:f+}
	\\
	\left(\rho \bar{v}_\phi\right) (R, z) \equiv & \int_{|\vec{v}| \leq \sqrt{2 \Psi(R, z)}} \mathrm{d}^3 v \; |\vec{v}| \cdot f(\mathcal{E}, L_z) = \frac{2 \pi}{R^2} \int_{0}^{\Psi(R, z)} \mathrm{d} \mathcal{E} \int_{-R \sqrt{2(\Psi(R, z) - \mathcal{E})}}^{R \sqrt{2(\Psi(R, z) - \mathcal{E})}} \mathrm{d}L_z \; L_z \cdot f_-(\mathcal{E}, L_z) \label {eqn:f-} \;.
\end{align}
In the formulas above $\bar{v}_\phi$ is the rotational velocity around the symmetry axis, while $R$ and $z$ are the radial distance and vertical height in the usual cylindrical coordinate frame.

Analogously to Eddington's formula, the two-integral-of-motion PSDFs can be reconstructed via an inversion of Eqs.~(\ref{eqn:f+}) and (\ref{eqn:f-}). In this work we follow the approach proposed by Hunter \& Qian~\cite{HQ1993}, which we refer in the following as ``HQ method". One starts with the analytic continuation of the density $\rho(R, z)$ and relative gravitational potential $\Psi(R, z)$ into the complex plane; restricting to models which are symmetric with respect to the equatorial plane, and provided that $\Psi$ decreases monotonically with increasing $z$, one can replace the cylindrical coordinates $R$ and $z$ with the variables $R^2$ and $\Psi$. After a few steps (see~\cite{HQ1993} for details), one can show that the $L_z$-even part of PSDF can be computed as:
\begin{align} \label{eqn:psdf}
f_+(\mathcal{E}, L_z) = \frac{1}{4 \pi^2 i \sqrt{2}} \oint_{C(\mathcal{E})} \frac{\mathrm{d} \xi}{\sqrt{\xi - \mathcal{E}}} \left. \frac{\mathrm{d}^2 \rho(R^2, \Psi)}{\mathrm{d} \Psi^2} \right|_{\substack{\Psi = \xi \;\;\;\;\;\;\;\;\;\; \\ R^2=\frac{L_z^2}{2(\xi - \mathcal{E})}}} \; ,
\end{align}	
where $C(\mathcal{E})$ is an appropriate path which tightly wraps around the real axis between the value of the potential at infinity, $\Psi_\infty$, and a characteristic value $\Psi_{\mathrm{env}}$, which, for any given value of the relative energy $\mathcal{E}$, is the value of the relative potential corresponding to the position on the galactic plane at which a circular orbit of radius $R_c$ has relative energy $\mathcal{E}$, namely $\Psi_{\mathrm{env}}(\mathcal{E}) = \Psi\left(R=R_c(\mathcal{E}),z=0\right)$. Hunter \& Qian propose, as useful way to parameterize the contour, to define an ellipse in terms of a real variable $\theta \in [0, 2 \pi]$:
\begin{align}
\xi (\theta) = \frac{\Psi_{\mathrm{env}}(\mathcal{E})}{2} (1 + \cos \theta) + i \, h \sin \theta \label{eqn:contour_finite} \; , \\
\xi (\theta) = \Psi_{\mathrm{env}}(\mathcal{E}) + l (1 - \sec \frac{\theta}{2}) + i \, h \sin \theta \label{eqn:contour_infinite} \; ,
\end{align}
where the first expression should be used in case of finite $\Psi_\infty$ and the second one in case of $\Psi_\infty \rightarrow -\infty$. The parameter $h$ controls the width of the contour in the imaginary plane, while $l$ is relevant only for infinite potentials and determines where the contour reaches its maximum width. In practice it is good to keep $h$ small to avoid including possible additional singularities that arise from the analytical continuation of $\rho$, however large enough to maintain good numerical convergence. Having this, the crucial point becomes the evaluation of the second derivative of the density with respect to the potential. In most cases one cannot perform the change of variables explicitly, and is forced to use the implicit derivation in cylindrical coordinates:
\begin{align} \label{eqn:derivative_expansion}
\frac{\mathrm{d}^2 \rho(R^2, \Psi)}{\mathrm{d} \Psi^2} = & \frac{\mathrm{d}^2 \rho(R^2, z^2)}{\mathrm{d} (z^2)^2} \left(\frac{\mathrm{d} \Psi(R^2, z^2)}{\mathrm{d} z^2} \right)^{-2} 
 - \frac{\mathrm{d} \rho(R^2, z^2)}{\mathrm{d} z^2} \frac{\Psi(R^2, z^2)}{\mathrm{d} (z^2)^2}\left(\frac{\mathrm{d} \Psi(R^2, z^2)}{\mathrm{d} z^2} \right)^{-3} \; ,
\end{align}
evaluated at $R^2=\frac{L^2}{2(\xi - \mathcal{E})}$ and $z^2$ such that $\Psi(R^2, z^2) = \xi$. Values of $z^2$ fulfilling the latter equality typically need to be found via numerical minimization routines. Further difficulties might arise if $\Psi(R^2,z^2)$ contains a branch cut along the contour, inducing a discontinuity in the Jacobian of the coordinate transformation; this typically happens for certain values of $\mathcal{E}$ and $L_z$ for a system embedded in very flattened potential and requires a proper adjustment of $C(\mathcal{E})$ and of the method in which the numerical integral is performed (we discuss this technical issue and its possible solutions in Appendix \ref{app:branch_cuts}). Finally, one can simplify the contour integral by using the Schwarz reflection principle, which implies that the values of integral above and below x-axis must be complex conjugates of each other. Therefore one can shrink the domain of integration to $\theta \in [0, \pi]$, compute only the real part and multiply the final result by factor of 2. The $L_z$-odd part of PSDF can be computed analogously, using the following expression:
\begin{align} \label{eqn:psdf_odd}
	f_-(\mathcal{E}, L_z) = \frac{\textrm{sign}(L_z)}{8 \pi^2 i} \oint_{C(\mathcal{E})} \frac{\mathrm{d} \xi}{\xi - \mathcal{E}} \left. \frac{\mathrm{d}^2 \left( \rho \bar{v}_\phi \right)}{\mathrm{d} \Psi^2} \right|_{\substack{\Psi = \xi \;\;\;\;\;\;\;\;\;\; \\ R^2=\frac{L_z^2}{2(\xi - \mathcal{E})}}} \; .
\end{align}
It is important to note that in order to evaluate $f_-$ one needs to specify also $\bar{v}_\phi(R^2,z^2)$, which is unfortunately often unknown. To surmount this one can either assume a parametric form for $\bar{v}_\phi$ or construct the PSDF using only $f_+$. We address this issue in greater detail in the following section.

In principle the HQ method can be used to compute the PSDF for any choice of axisymmetric $\rho(R^2, z^2)$, $\Psi(R^2, z^2)$ and $\bar{v}_\phi(R^2, z^2)$, however there is no guarantee that the resulting PSDF will be positive definite (i.e. physical). This needs to be checked explicitly after performing the contour integrals. At this point we also note that one can check the accuracy of resulting PSDF by, for example, plugging it back in Equation~\eqref{eqn:f+}, which should reproduce the initially assumed density distribution. 
In our analysis we were able to reconstruct the initial density within one percent accuracy in the regions of interest for all the studied cases.

\subsection{Modeling of spiral galaxies}
\label{sec:spirals_model}

The HQ method, described in the previous section, turns out to be indispensable for determining the phase-space distributions within spiral galaxies. The method can be applied to the DM halo, but also stars or any other component that is well approximated by steady-state distribution of collisionless point-like objects. To obtain $f_+$ for the component of interest, one needs to specify its spatial distribution, as well as the total gravitational potential. Spiral galaxies are typically composed of a stellar disc with a bulge/bar structure in the center, embedded in large DM halo. We consider a toy model with: {\sl i)} a Myamoto-Nagai (MN) disc, with potential:
\begin{align} \label{eqn:potential_MN}
	\Psi\sub{MN}(R^2, z^2) = \frac{G M_d}{\sqrt{R^2 + (a_d + \sqrt{z^2 + b_d^2})^2}}
\end{align}
parametrized by the mass $M_d$, the characteristic radius $a_d$ and the characteristic height $b_d$; {\sl ii)} a spherically symmetric  Hernquist bulge (we are not going to discuss results regarding regions where the bulge is the dominant component, hence this specific choice is not crucial), 
with potential:
\begin{align}
	\Psi\sub{Her}(R^2, z^2) = \frac{G M_b}{\sqrt{R^2 + z^2} + a_b}
\end{align}
parametrized by its mass $M_b$ and characteristic radius $a_b$; and {\sl iii)} a spheroidal DM halo with a NFW density profile~\cite{Navarro:1996gj} (again, we are not going to zoom to the very central region of spiral galaxies, so results we are presenting are not crucially dependent on this specific choice):
\begin{align} \label{eqn:nfw}
	\rho\sub{NFW}(m) = \frac{\rho_s}{m / r_s \cdot (1 + m / r_s)^2} \;\;\; \textrm{where} \;\;\; m^2 = R^2 + z^2 / q^2 \; ,
\end{align}
parametrized by the scale density $\rho_s$, the scale radius $r_s$ and the ``flattening" parameter $q$.
For spherical halos, obtained by setting $q = 1$, the corresponding gravitational potential can be computed analytically:
\begin{align}
	\Psi\sub{NFW}(r) = 2 \pi G \rho_s r_s^2 \cdot \frac{\log(1 + r / r_s))}{r / r_s} \;\;\; \textrm{where} \;\;\; r = \sqrt{R^2 + z^2} \; ,
\end{align}
while for oblate ($q < 1$) or prolate ($q > 1$) halos, a numerical evaluation of the following integral is required:
\begin{align}
	\Psi\sub{NFW}(R^2, z^2) = \pi G q \int_0^\infty \frac{\mathrm{d}u}{(1 + u)\sqrt{q^2 + u}} \int_U^\infty \rho(m^2) \mathrm{d}m^2 
	\quad \quad \textrm{where} \quad \quad U = \frac{R^2}{1 + u} + \frac{z^2}{q^2 + u} \, . 
\end{align}
The above model involves a set of free parameters that need to be inferred from observations. In this section (unless specified otherwise), we will mostly refer to a sample case in which the number of free parameters is reduced introducing the following correlations, which are in rough agreement with what is typically found in spiral galaxies~\cite{2007MNRAS.378...41S,Courteau2015,Soufe2016}:
\begin{align}
	M_b = 0.05 M\sub{2.2} \;\;\; , \;\;\; M_d = 0.45 M\sub{2.2} \\
	a_b = \frac{a_d}{3} \;\;\; , \;\;\; b_d = \frac{a_d}{10} \;\;\; , \;\;\; r_s = 5 a_d
\end{align}
where $M\sub{2.2}$ is the total mass of the object within a radius equal to $2.2$ disc lengths $a_d$. This characteristic scale turned out to be particularly useful as a benchmark distance for determining the fraction of DM mass in a given galaxy~\cite{Courteau1999}. We use the corresponding circular velocity $\hat{V}_c \equiv V_c(R=2.2 a_d)$ as a normalization scale in the rest of this section.

\subsection{From spherical to axial symmetry}

\begin{figure}
	\includegraphics[width=0.49\textwidth]{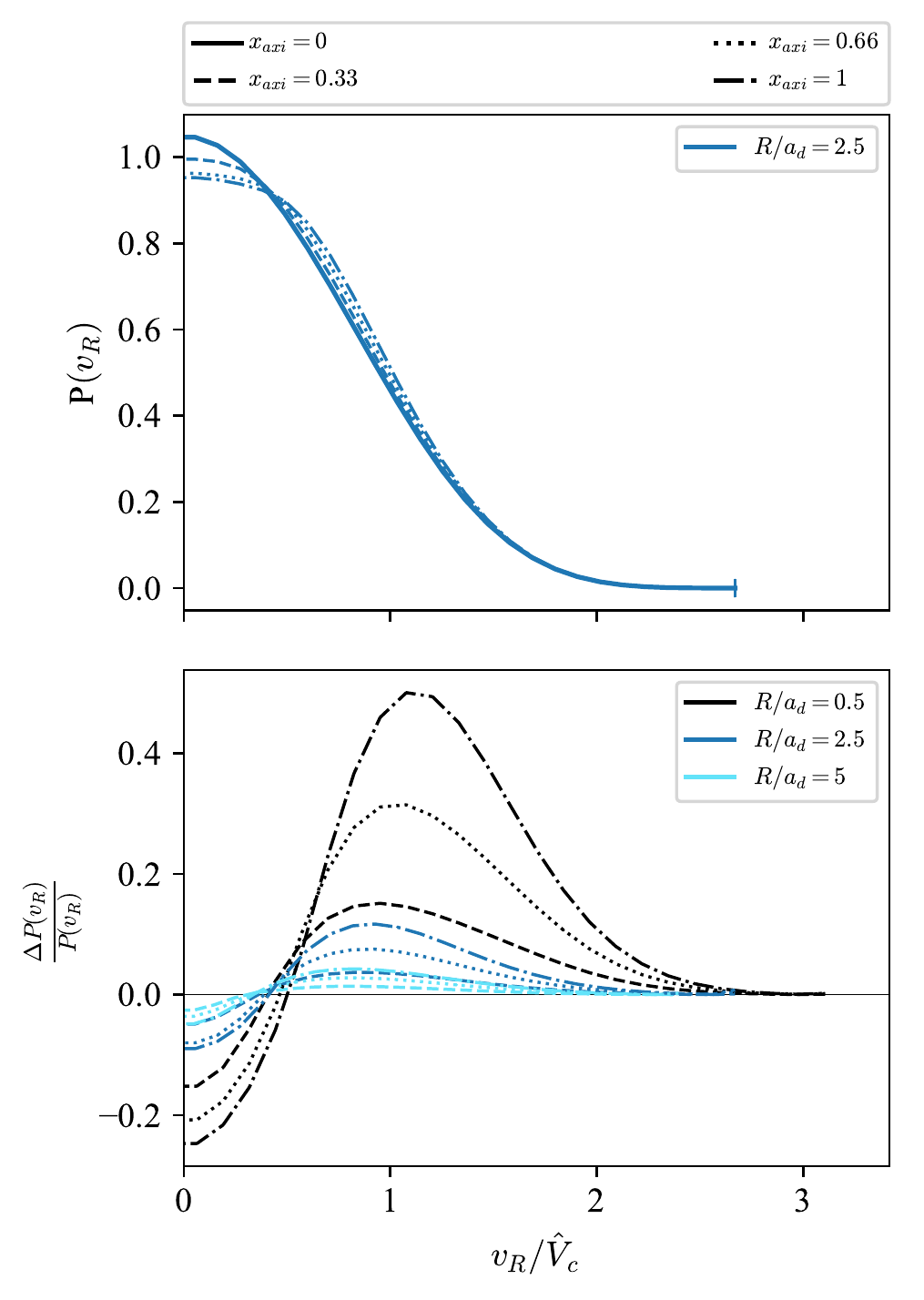}
	\includegraphics[width=0.49\textwidth]{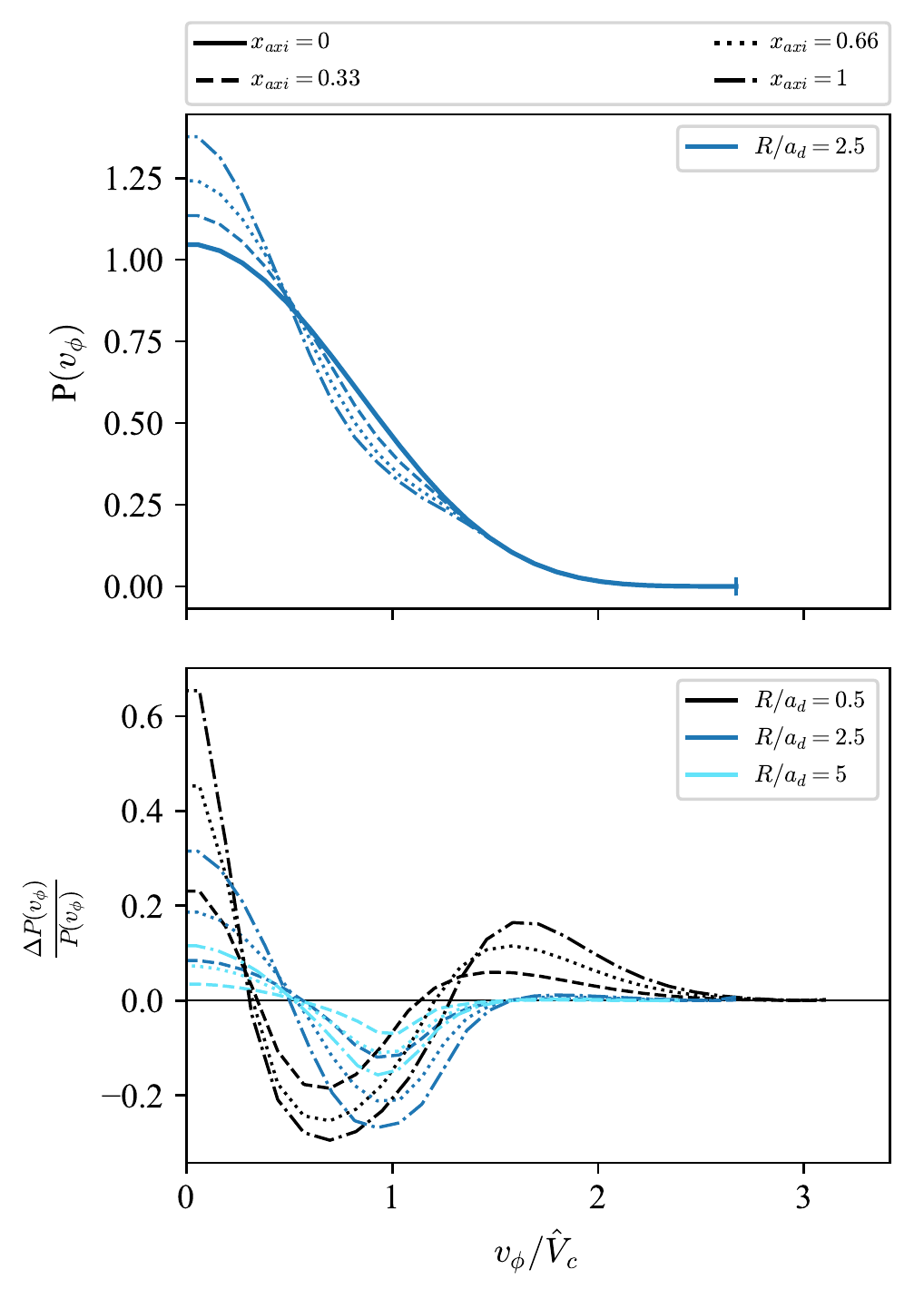}
	\caption{Radial (left) and azimuthal (right) velocity distributions in the galactic plane for various fraction of the axisymmetric component, parametrized by $x\sub{axi}$. In lower panels we show the relative difference with respect to the $x\sub{axi}=0$ case, computed at different radii.}
	\label{fig:sphericity}
\end{figure}

\begin{figure}
	\includegraphics[width=0.49\textwidth]{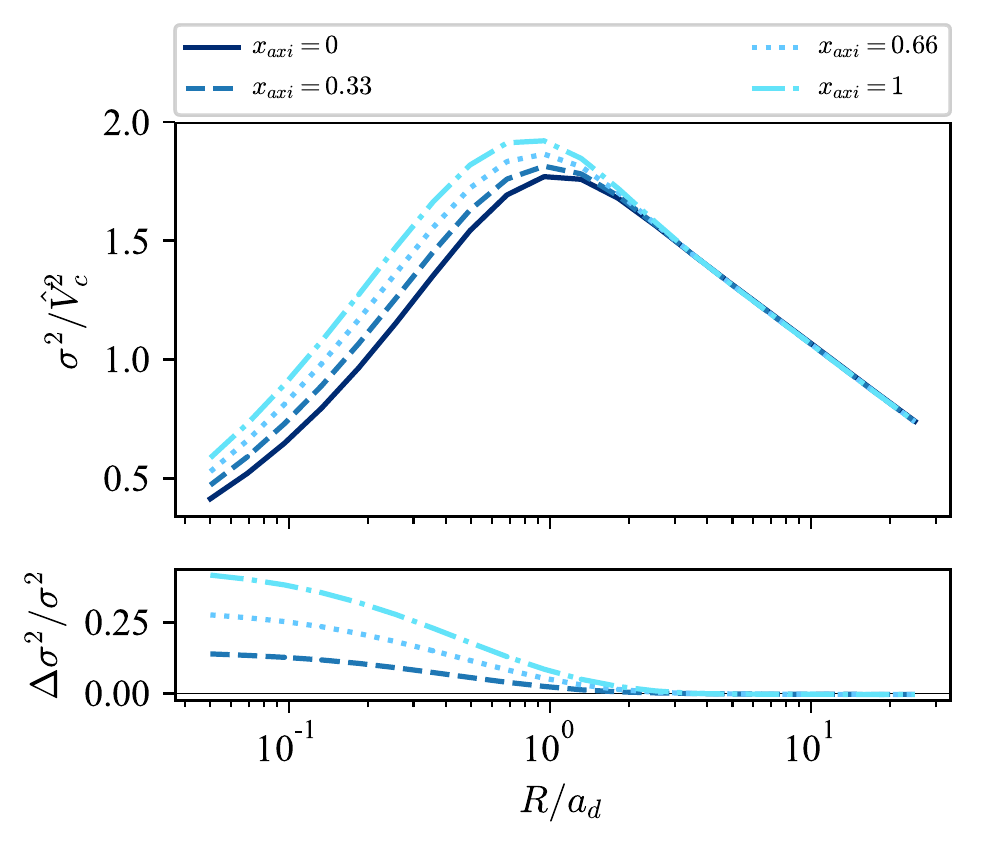}
	\includegraphics[width=0.49\textwidth]{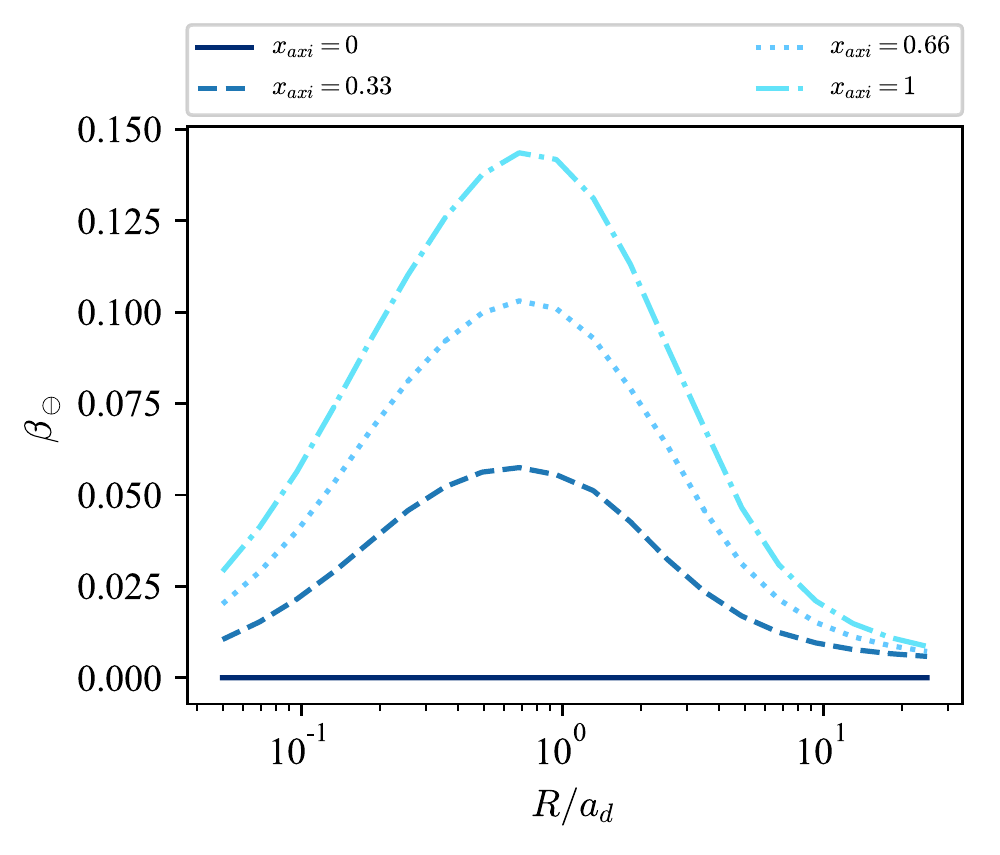}
	\caption{Total velocity dispersion (left) and velocity anisotropy (right) in the galactic plane as a function of radial distance for various fractions of axisymmetric component, parametrized by $x\sub{axi}$. In the lower left panel we show the relative difference of the velocity dispersion with respect to the $x\sub{axi}=0$ case.}
	\label{fig:sphericity_disp}
\end{figure}

When connecting the model to observations, one needs to fit the total gravitational potential $\Psi\sub{tot}$, which is the sum of bulge, disc and halo components, to reproduce the observed circular velocity profile in the galactic plane $V_c(R)$:
\begin{align}
	V^2_c(R) = - 2 R^2 \cdot \left. \frac{\textrm{d} \Psi\sub{tot}(R^2, z^2)}{\mathrm{d}R^2} \right|_{z=0} \; .
\end{align}
It is evident that spherical, as well axysimmetric, modeling can reproduce given $V_c(R)$. Hence, spherically symmetric models were used in the past to compute PSDF of halos using the Eddingtons inversion. We demonstrate that such simplification can drastically affect the PSDF and the axisymmetric HQ method should be used instead. To illustrate the difference, we used a combination of spherical NFW potential for the halo and a linear combination of MN and Plummer potential (spherical approximation of MN potential, $\Psi\sub{Plu}(r^2) \sim \Psi\sub{MN}(R^2 \rightarrow r^2, z^2 \rightarrow 0)$) for the disc:
\begin{align}
	\Psi\sub{disc}(R^2, z^2) = x\sub{axi} \Psi\sub{MN}(R^2, z^2) + (1 - x\sub{axi}) \Psi\sub{Plu}(R^2+z^2) 
\end{align}
(for the moment we omit the sub-dominant bulge component). In Figure~\ref{fig:sphericity} we present the comparison of radial and azimuthal velocity distributions, as well as the residuals with respect to the spherical limit, for various values of $x\sub{axi}$. As the admixture of the axisymmetic potential increases, the radial velocity distribution becomes shifted towards higher velocities, while the azimuthal component gains power at low velocities. This results in skewed velocity distributions that can not be accurately modelled within the spherical approximation, nor using a Gaussian profile. The differences are most significant in central part of the halo and gradually diminish with increasing distance from the center, as the effect of disc component becomes negligible. The changes in the velocity distributions naturally lead also to changes in the velocity dispersion $\sigma^2(R,z)$, as well as in a velocity anisotropy. Analogously to spherical systems, for which the anisotropy is usually described in terms of an anisotropy parameter defined as ratio of velocity second moments in tangential and radial direction, we introduce here the following quantity, better suited for describing axial systems:
\begin{align} \label{eqn:anisotropy}
\beta_\ominus (R,z) \equiv \frac{1}{2} - \frac{\sigma^2_\phi(R^2, z^2)}{\sigma^2_M(R^2, z^2)} \; ,
\end{align}
where $\sigma^2_\phi$ is the velocity dispersion in azimuthal direction, while $\sigma^2_M = \sigma^2_R + \sigma^2_z$ is the velocity dispersion in meridional plane (note that for $f(\mathcal{E},L_z)$ the velocity dispersion in meridional plane is isotropic, i.e. $\sigma^2_R = \sigma^2_z$, by construction). On the right hand side of Figure~\ref{fig:sphericity_disp} we show that the velocity anisotropy in the galactic plane, $\beta_\ominus(R,0)$, becomes increasingly radial as the admixture of axisymmetric potential increases, with the radial velocity dispersion increasing with $x\sub{axi}$ and the azimuthal component diminishing. In the plot on left hand side of same Figure~\ref{fig:sphericity_disp}, we show the total velocity dispersion ($\sigma^2 = 2 \sigma_R^2 + \sigma_\phi^2$), which also increases with $x\sub{axi}$. These effects are, however, again limited only to the central part of the galaxy, where the influence of disc is significant, and slowly diminish as one moves towards the outskirts of the system.

\begin{figure}
	\includegraphics[width=0.48\textwidth]{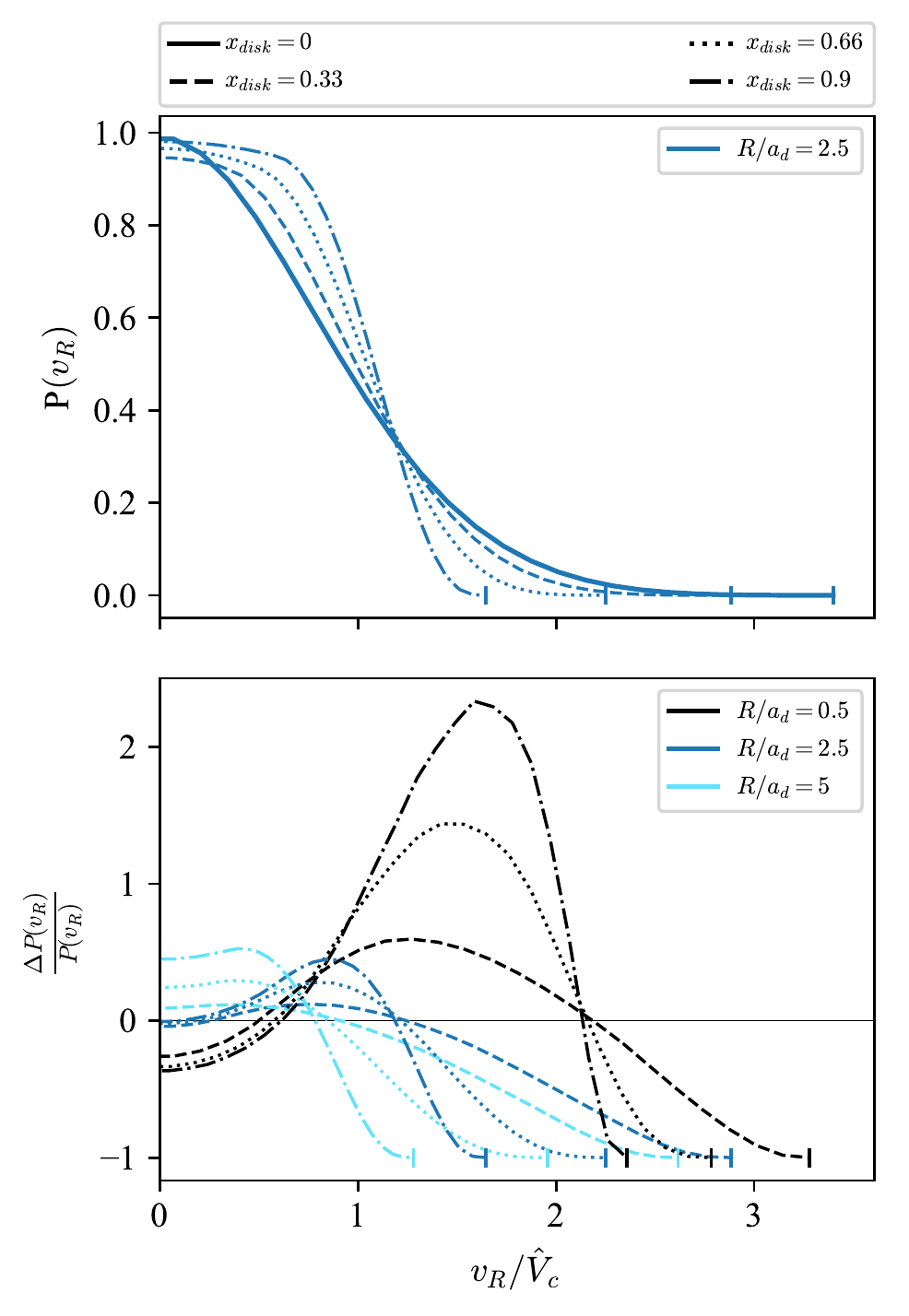}
	\includegraphics[width=0.50\textwidth]{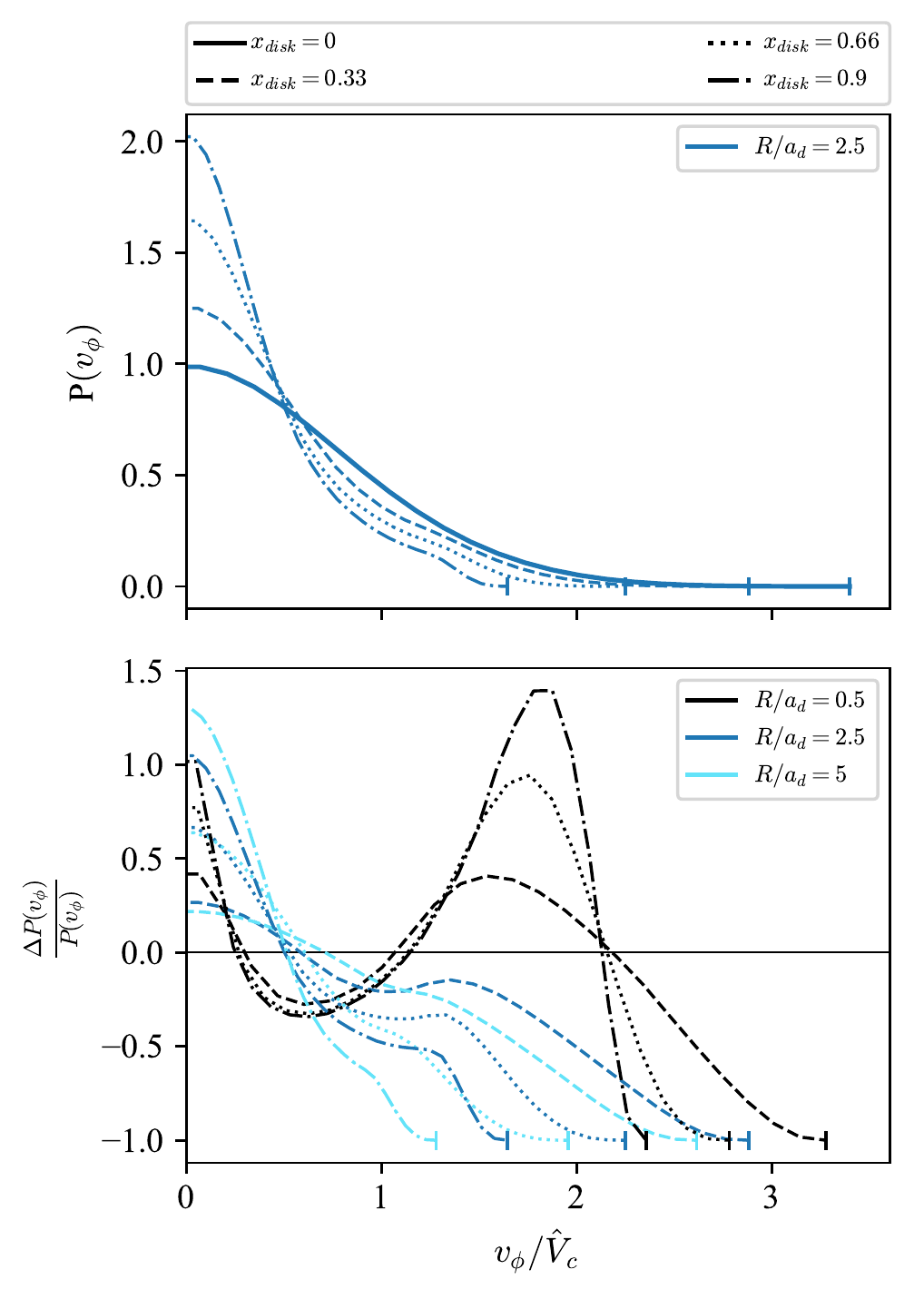}
	\caption{Radial (left) and azimuthal (right) velocity distributions in the galactic plane for various fractions of the stellar disc component, parametrized by $x\sub{disc}$. In the lower panels we show the relative difference with respect to the $x\sub{disc}=0$ case, computed at different radii.}
	\label{fig:baryons}
\end{figure}

Similarly, one can check how the PSDF of the halo particles changes if one varies the relative weight of axisymmetric disc and spherical halo components, while keeping the characteristic circular velocity $\hat{V}_c$ unchanged:
\begin{align}
	\hat{V}_c = - 2 \cdot (2.2 a_d)^2 \left( x\sub{disc} \left. \cdot \frac{\textrm{d} \Psi\sub{MN}(R^2, z^2)}{\mathrm{d}R^2} \right|_{\substack{R=2.2 a_d \\ z=0 \;\;\;\;\;\;\;}} + (1 - x\sub{disc}) \cdot \left. \frac{\textrm{d} \Psi\sub{NFW}(R^2, z^2)}{\mathrm{d}R^2} \right|_{\substack{R=2.2 a_d \\ z=0 \;\;\;\;\;\;\;}} \right)
\end{align}
In Figure~\ref{fig:baryons} we show the comparison of radial and azimuthal velocity distributions for various values of $x\sub{disc}$. The span of values displayed go from $x\sub{disc}=0$ (when the flattening induced by the disc is ignored), to values that are representative of minimal/average/maximal disc models for spiral galaxies. The trends are similar as when varying $x\sub{axi}$, since also $x\sub{disc}$ interpolates between spherically symmetric and axisymmetric configurations. However, an important difference is the fact that the local escape velocity, $v\sub{esc} = \sqrt{2 \Psi(R^2,z^2)}$, decreases with increasing $x\sub{disc}$, since smaller amount of total mass is needed to produce the same $\hat{V}_c$. Therefore the corresponding velocity distributions become suppressed at high velocities with respect to the halo-only case. This effect is somewhat compensated by the aforementioned shift of power in the radial velocity distribution towards higher $v$, which occurs in presence of flattened disc. The differences are again most significant in the central part of the galaxy, where the disc component dominates, but remain noticeable even at large radii due to the change in $v\sub{esc}$. The interplay of these effects again highlights the need for careful modelling that goes beyond the standard approximations.

\begin{figure}
	\includegraphics[width=0.49\textwidth]{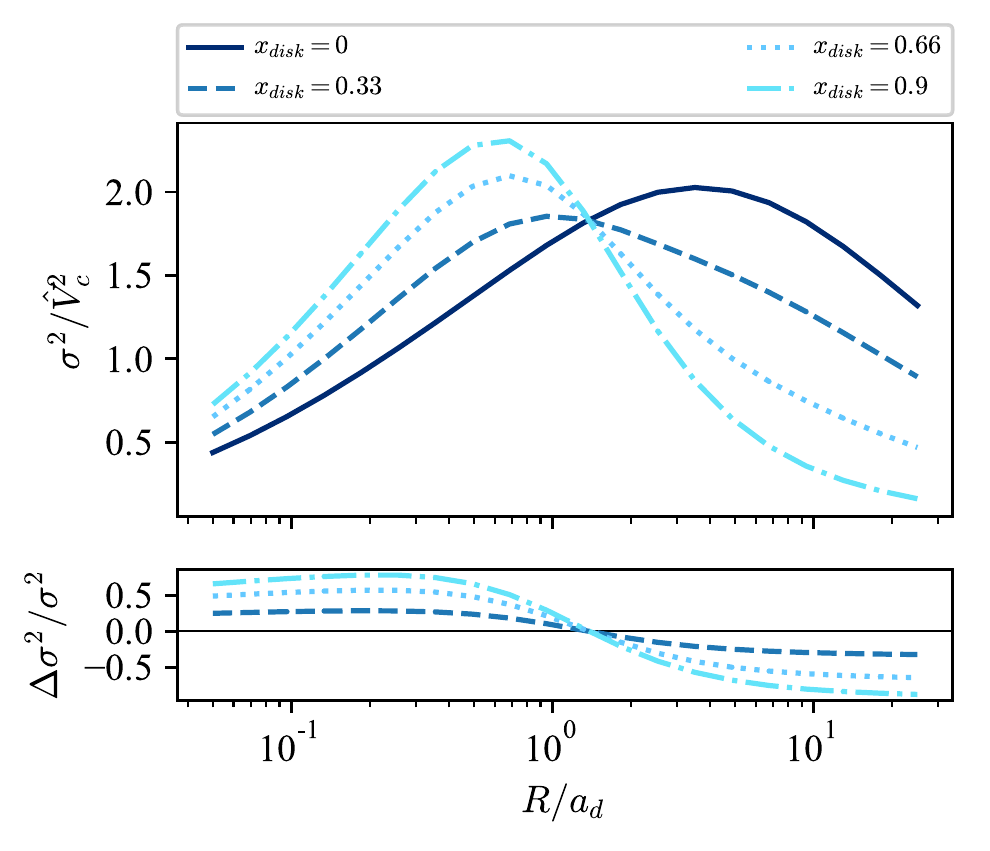}
	\includegraphics[width=0.49\textwidth]{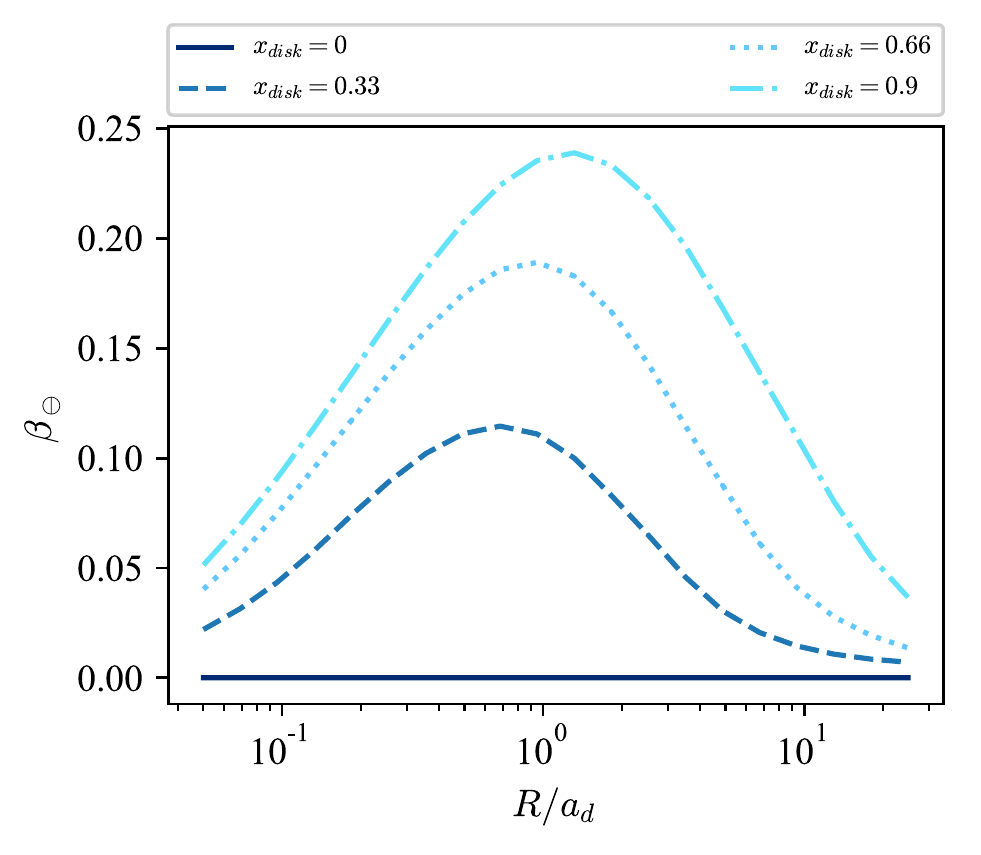}
	\caption{Total velocity dispersion (left) and velocity anisotropy (right) in the galactic plane as a function of radial distance for various fractions of the stellar disc component, parametrized by $x\sub{disk}$. In the lower left panel we show the relative difference of velocity dispersion with respect to the $x\sub{disc}=0$ case.}
	\label{fig:baryons_disp}
\end{figure}

\begin{figure}
	\includegraphics[width=0.49\textwidth]{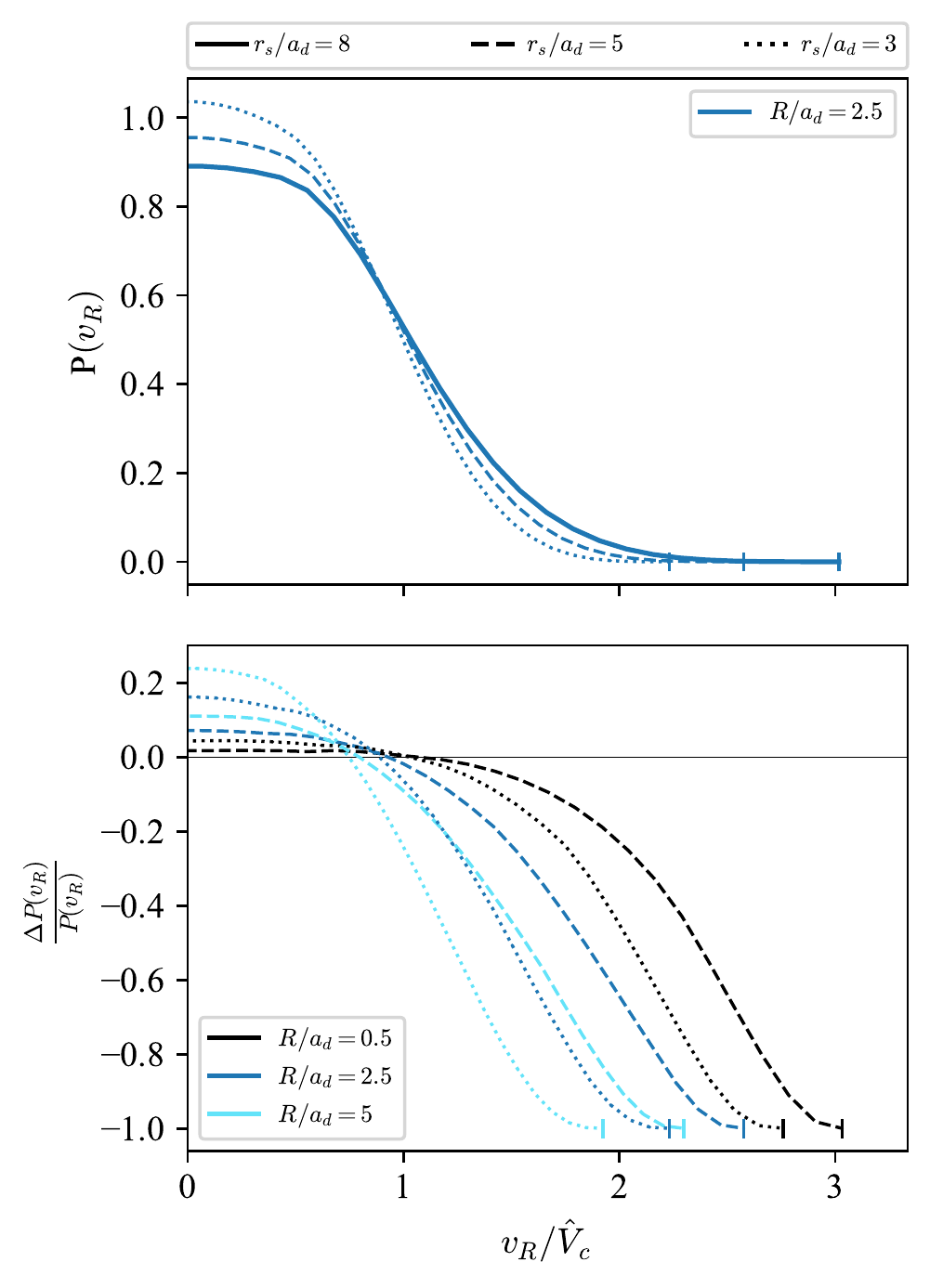}
	\includegraphics[width=0.50\textwidth]{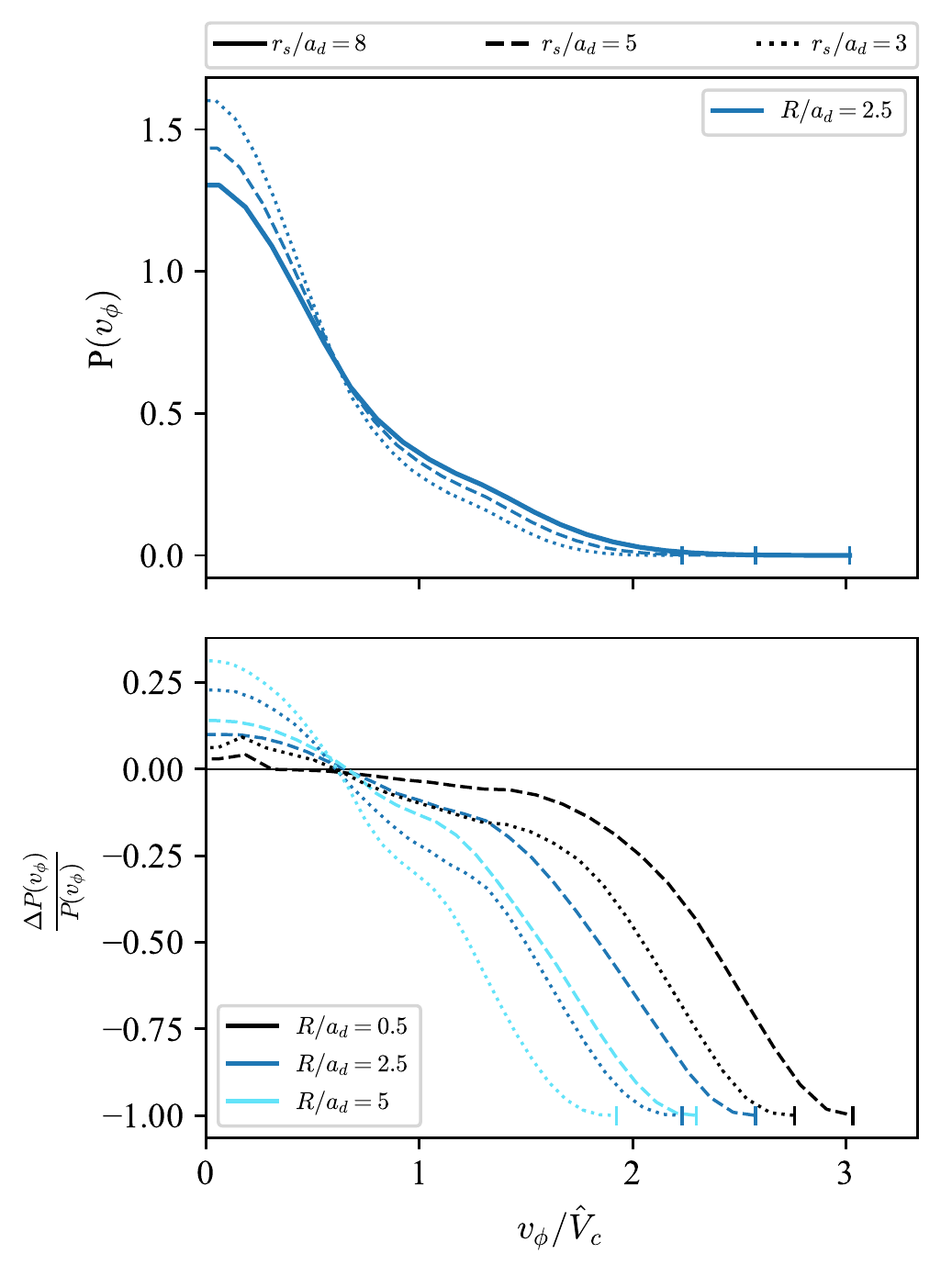}
	\caption{Radial (left) and azimuthal (right) velocity distributions in the galactic plane for various ratios of $r_s/a_d$. In the lower panels we show the relative difference with respect to the $r_s / a_d = 8$ case, computed at different radii.}
	\label{fig:rs}
\end{figure}

\begin{figure}
	\includegraphics[width=0.49\textwidth]{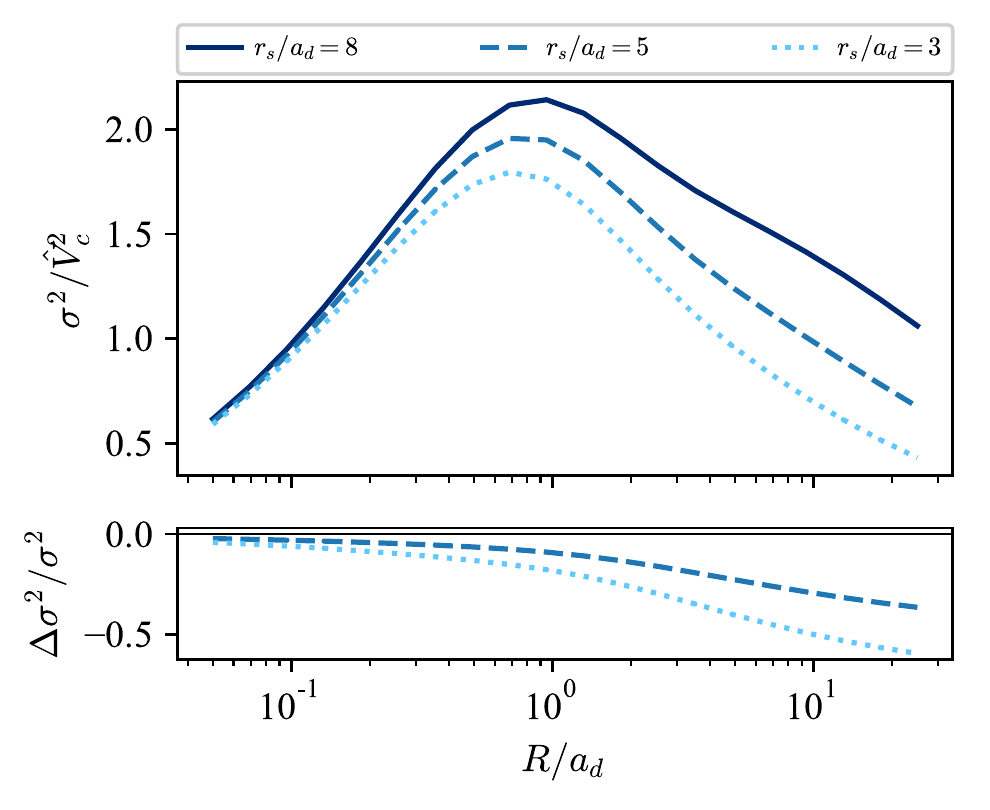}
	\includegraphics[width=0.49\textwidth]{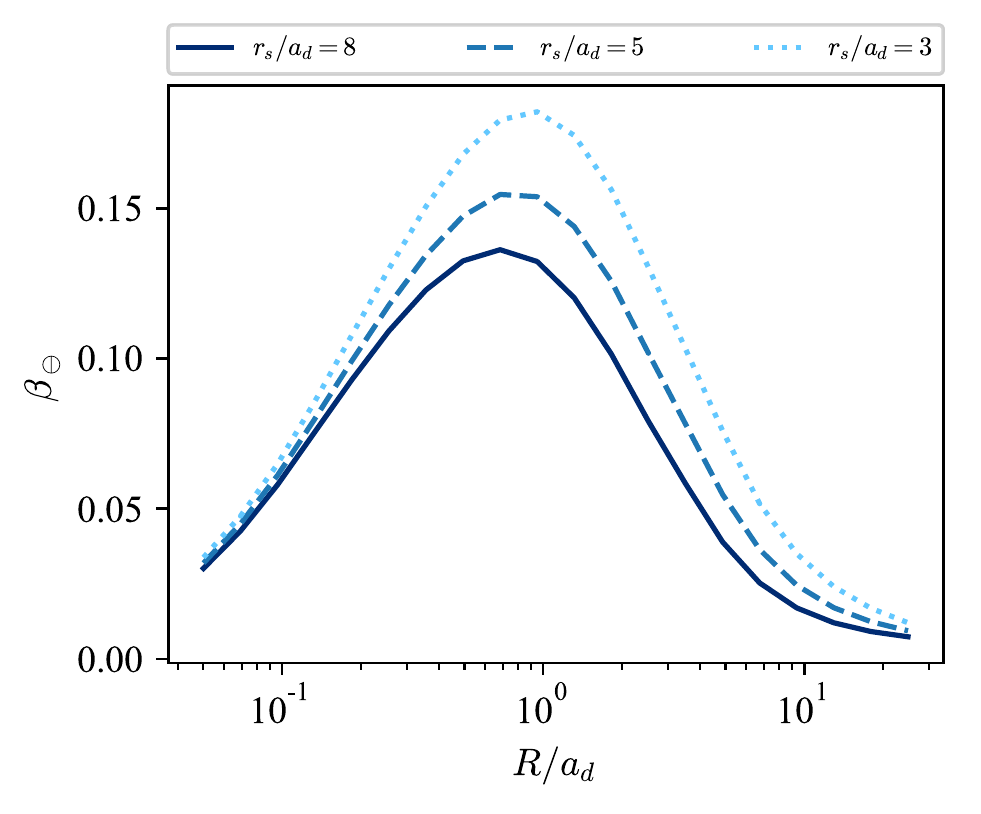}
	\caption{Total velocity dispersion (left) and velocity anisotropy (right) in the galactic plane as a function of radial distance for various ratios of $r_s/a_d$. In the lower left panel we show the relative difference of velocity dispersion with respect to the $r_s/a_d=8$ case.}
	\label{fig:rs_disp}
\end{figure}

Degeneracies in mass model decompositions sometimes make it hard to precisely infer the DM density profile. Beside the well known cusp/core problem, one often faces large uncertainties in inferring the halo scale radius $r_s$. In Figure~\ref{fig:rs} we show the velocity probability distributions for three different $r_s/a_d$ ratios, which cover a range of values typically encountered in disc galaxies~\cite{Soufe2016}. We find that for large $r_s / a_d$ the velocity distributions contains features that can not be encaptured by a simple Gaussian curve, while decreasing the ratio leads to increasingly Maxwellian distribution. An important difference with respect to varying the admixture of disc component is that both radial and azimuthal velocity distributions get shifted towards lower velocities as $r_s$ decreases, which leads to ``colder" halos at $R \gtrsim a_d$. This can also be seen from the corresponding velocity dispersion, shown in the plot on left hand side of Figure~\ref{fig:rs_disp}. In the right hand side plot of same Figure~\ref{fig:rs_disp}, we show the resulting velocity anisotropy in the galactic plane for same $r_s/a_d$ ratios, which demonstrates that decreasing $r_s$ make the DM particle trajectories increasing radially biased.

\begin{figure}
	\includegraphics[width=0.49\textwidth]{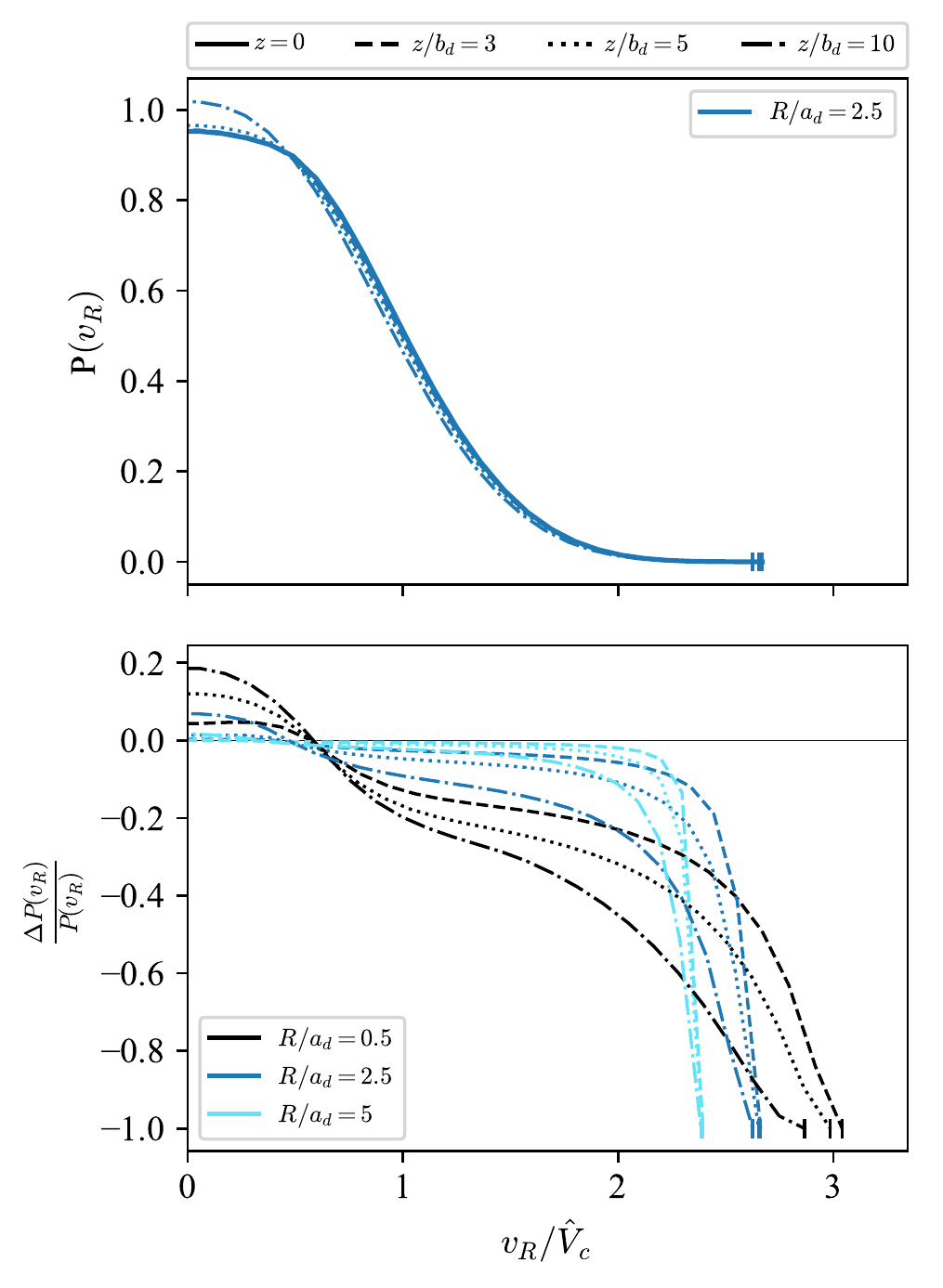}
	\includegraphics[width=0.50\textwidth]{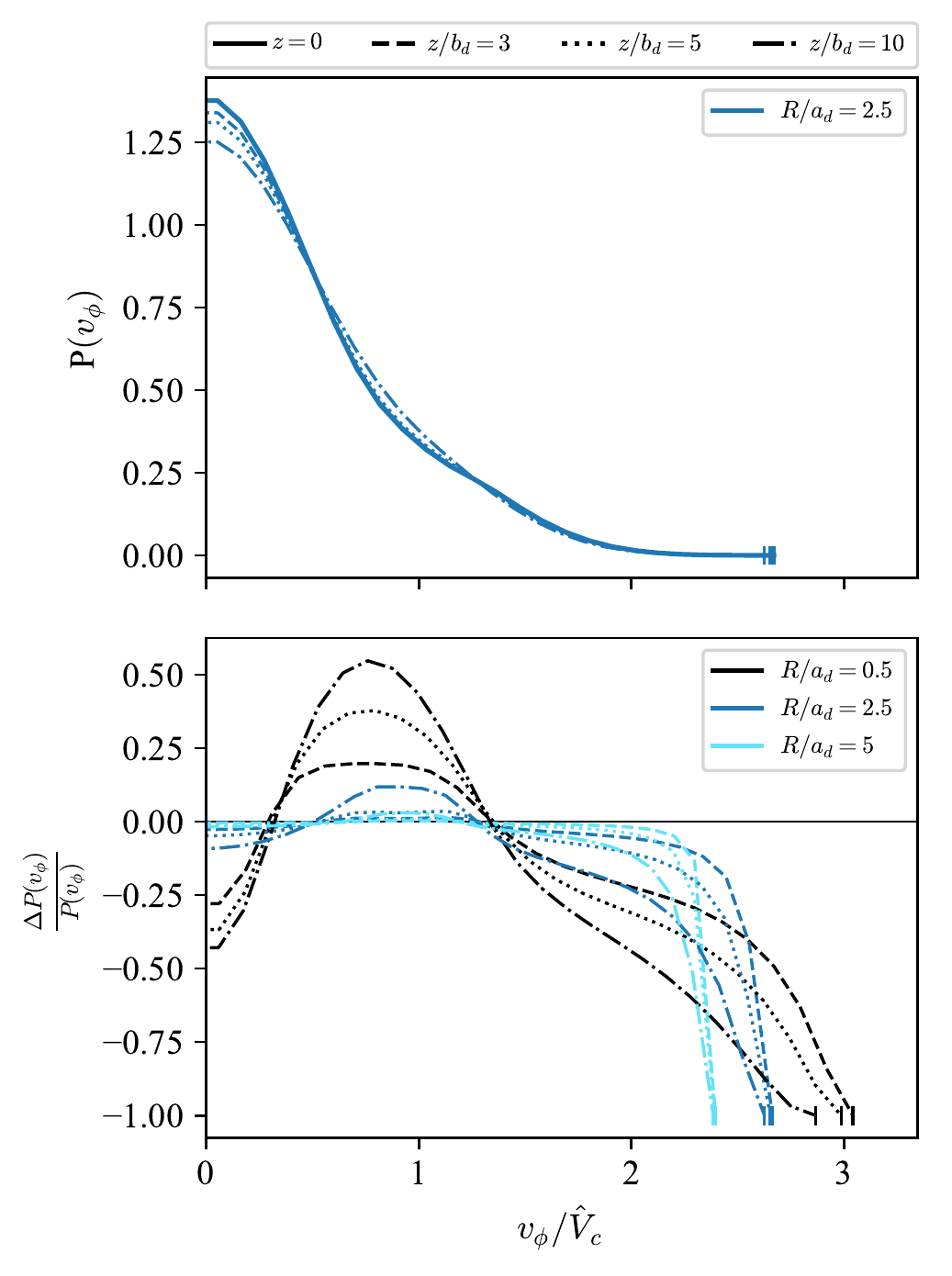}
	\caption{Radial (left) and azimuthal (right) velocity distributions for various heights $z$ above the galactic disc, normalized to the disc height $b_d$. In the lower panels we show the relative difference respect to the $z=0$ case, computed at different radii.}
	\label{fig:z}
\end{figure}

\begin{figure}
	\includegraphics[width=0.49\textwidth]{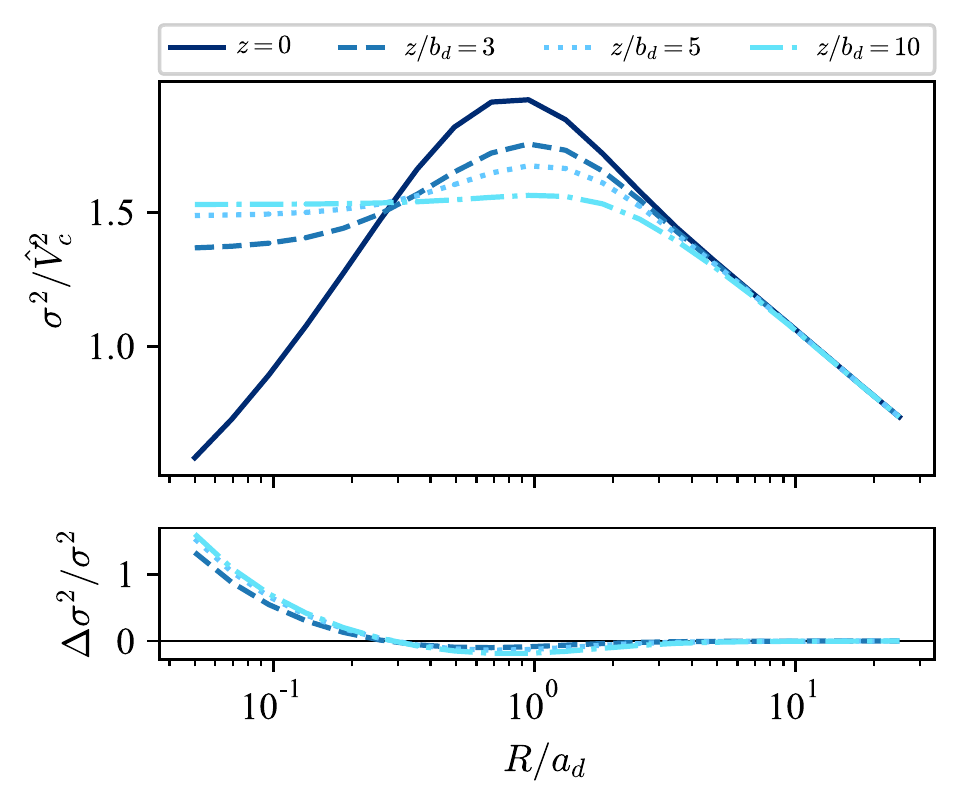}
	\includegraphics[width=0.49\textwidth]{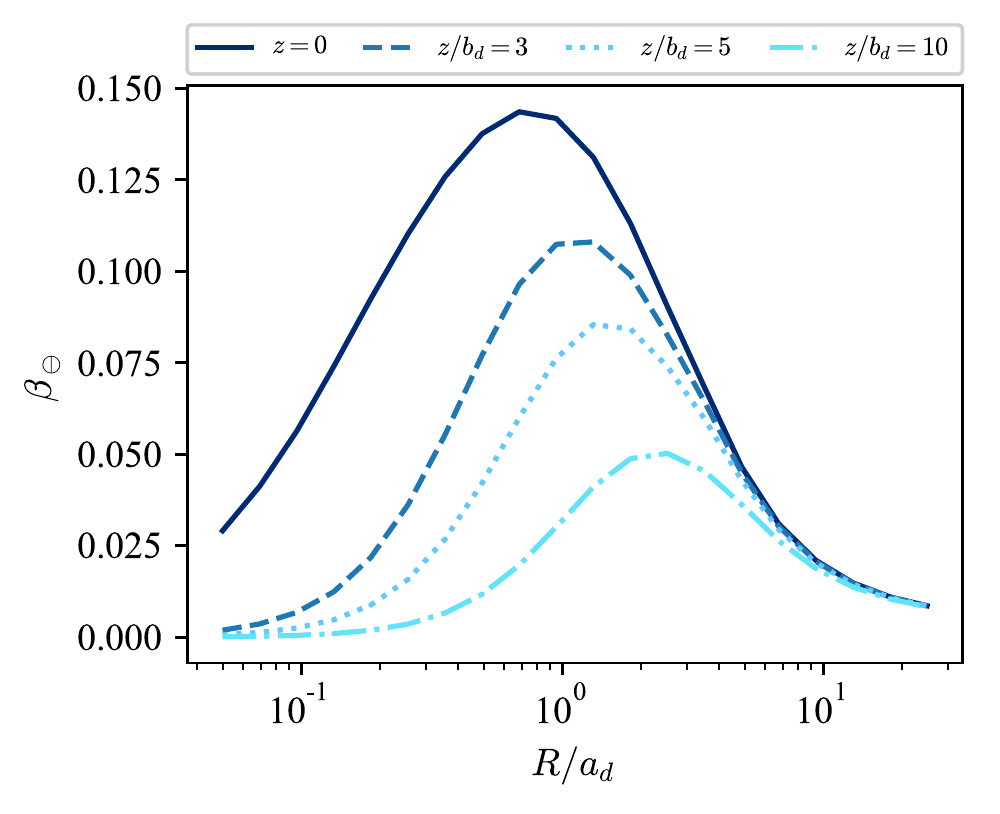}
	\caption{Total velocity dispersion (left) and velocity anisotropy (right) as a function of radial distance for various heights $z$ above the galactic disc (normalized to the disc height $b_d$). In the lower left panel we show the relative difference of velocity dispersion with respect to the $z=0$ case.}
	\label{fig:z_disp}
\end{figure}

Finally, we note that the effects of the stellar disc on the halo particles softens more rapidly as one moves along the $z$-axis, as compared to the radial direction. This can be seen from Figure~\ref{fig:z}, where we plot the radial and azimuthal velocity probability distributions for various heights above the galactic plane. As one moves towards larger values of $z$, the radial velocity distribution gets shifted back to lower velocities, while the azimuthal components gets more power at high velocities. These trends are the opposite as one finds for increasing the amount of disc component. Furthermore, at $z \gg b_d$ the velocity distribution becomes closer to isotropic, which is most significant for $R \lesssim a_d$, while the velocity dispersions are driven towards constant central value, as shown in Figure~\ref{fig:z_disp}.

\subsection{Halo flattening}

\begin{figure}
	\includegraphics[width=0.49\textwidth]{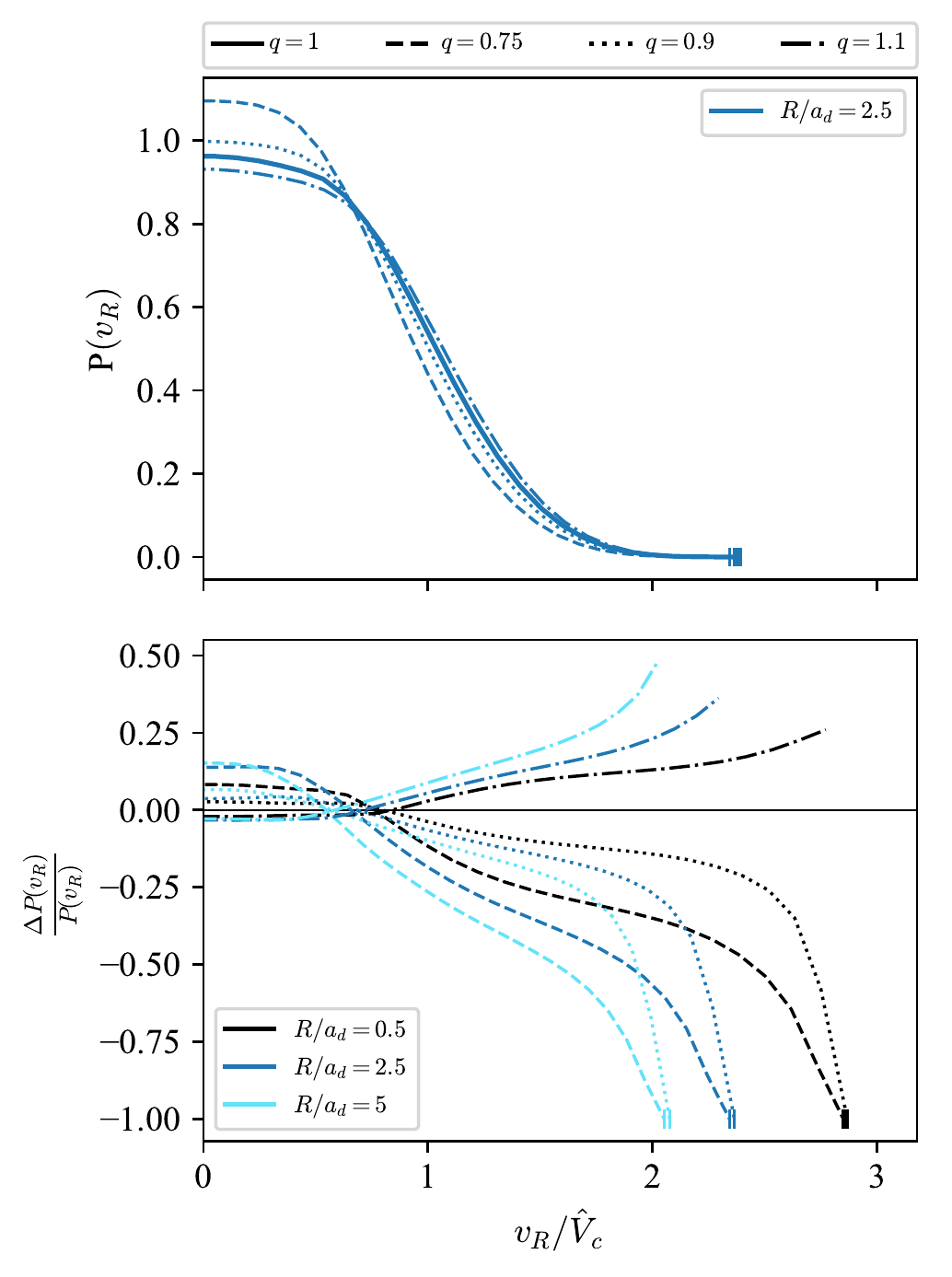}
	\includegraphics[width=0.49\textwidth]{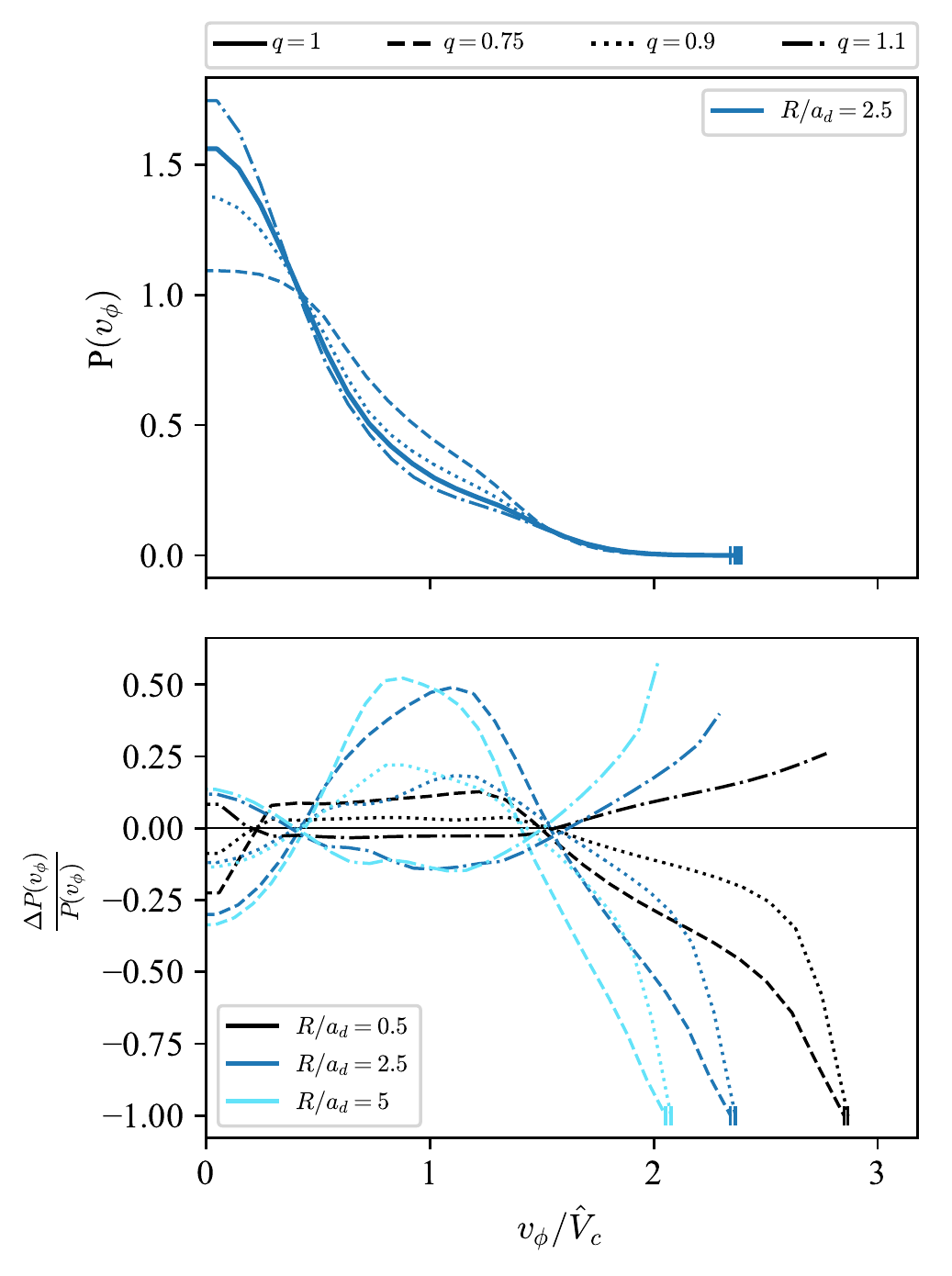}
	\caption{Radial (left) and azimuthal (right) velocity distributions in the galactic plane for various halo shapes, parametrized by $q$. In the lower panels we show the relative difference with respect to the spherical halo, computed at different radii.}
	\label{fig:q}
\end{figure}

\begin{figure}
	\includegraphics[width=0.49\textwidth]{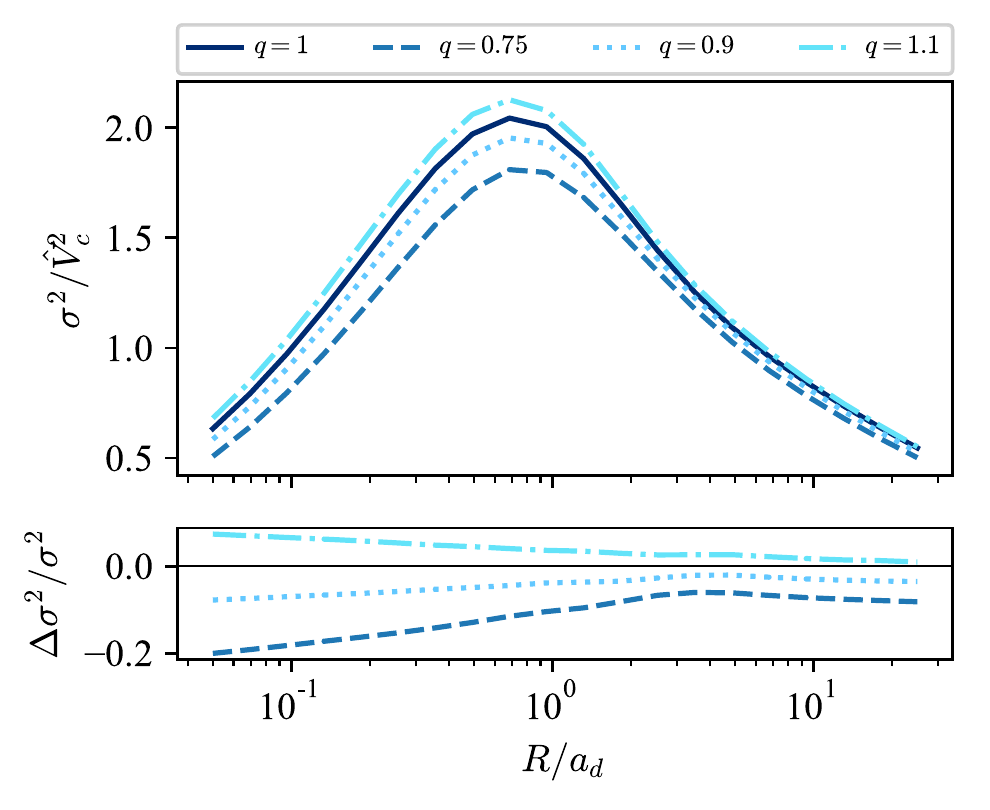}
	\includegraphics[width=0.49\textwidth]{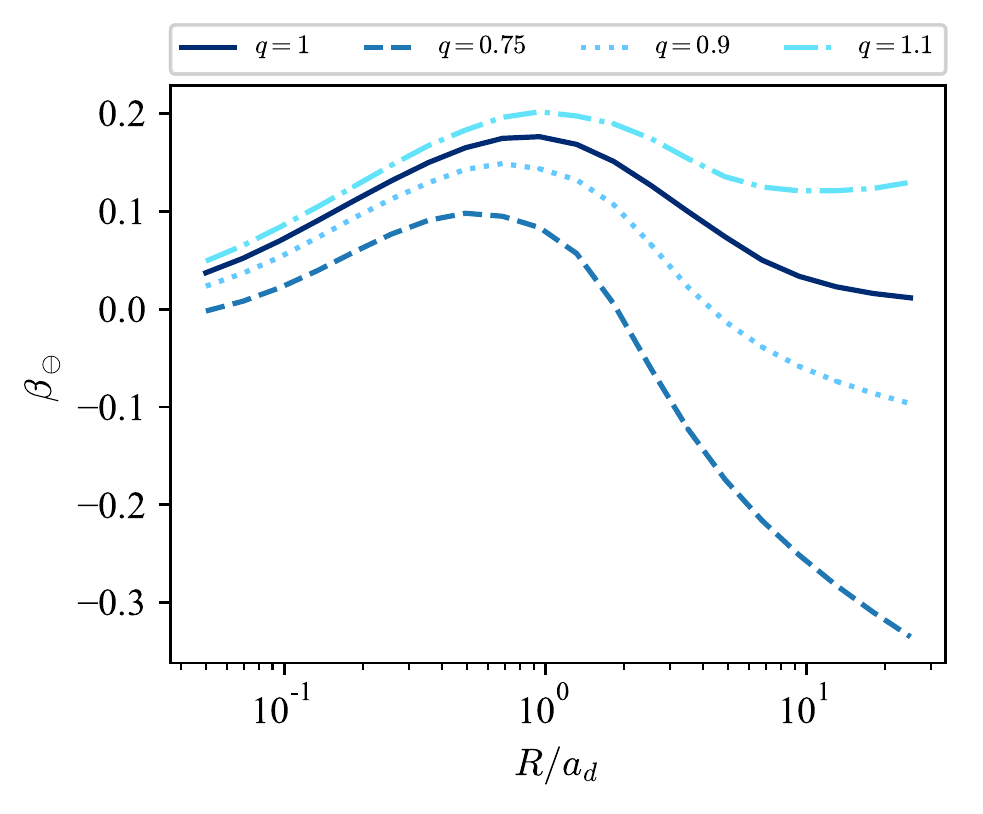}
	\caption{Total velocity dispersion (left) and velocity anisotropy (right) in the galactic plane as a function of radial distance for various halo shapes, parametrized by $q$. In the lower left panel we show the relative difference of velocity dispersion with respect to the spherical halo.}
	\label{fig:q_disp}
\end{figure}

Using the HQ method one can also study the effect of flattening or elongation of the DM halo along the axis of symmetry. As already mentioned, in our toy model the sphericity of halo is controlled by parameter $q$, where $q < 1$ corresponds to oblate and $q > 1$ to prolate configurations. Even though some recent studies suggest that the Milky Way halo is very close to spherical~\cite{Bovy2016:q}, there is substantial evidence for a highly flattened DM sub-component coming from past mergers, see~\cite{Evans:2018bqy} and references therein. Furthermore, hydrodynamical simulations of structure growth, which include baryons, generically predict oblate halos~\cite{Bryan2013,Chua2018} with significant spread in the $q$ parameter, whose value is strongly influenced by the formation history of the particular object. In Figure~\ref{fig:q} we show the halo velocity distributions obtained for a range of typical $q$ values, while keeping the baryonic component fixed. We find that for oblate halos the radial velocity distribution gets shifted towards smaller velocities, while the azimuthal velocity distribution is boosted at intermediate velocities and suppressed elsewhere. The morphology of the halo also affects the depth of the gravitational potential when keeping $\hat{V}_c$ fixed, which in turn leads to lower $v\sub{esc}$ for $q < 1$, while changes are in the opposite direction in case of prolate halos. By comparing with Figures~\ref{fig:sphericity} and~\ref{fig:baryons} we can see that flattening has roughly the opposite effect of increasing the disc component, however the velocity distributions in presence of both features are still poorly described by the Gaussian or spherical approximation (this is discussed in greater details in the next section, when considering the Milky Way). The corresponding velocity dispersion, portrayed in the left hand side plot of Figure~\ref{fig:q_disp}, is consistent with the changes in the velocity probability distributions, as it decreases (increases) for oblate (prolate) halos. The effect remains significant at all radii, since the halo is the largest component of galaxy and extends way beyond the stellar disc. The velocity anisotropy in the galactic plane, showed in the right hand side plot of Figure~\ref{fig:q_disp}, remains radially biased in the central part for oblate and prolate halos, however in the outskirts we see different behaviors. Oblate halos generically lead to negative, i.e. circularly biased, velocity anisotropy, while for prolate halos we find increasing radial bias at large galactocentric distances.

\subsection{Halo rotation}
	
\begin{figure}
	\includegraphics[width=0.48\textwidth]{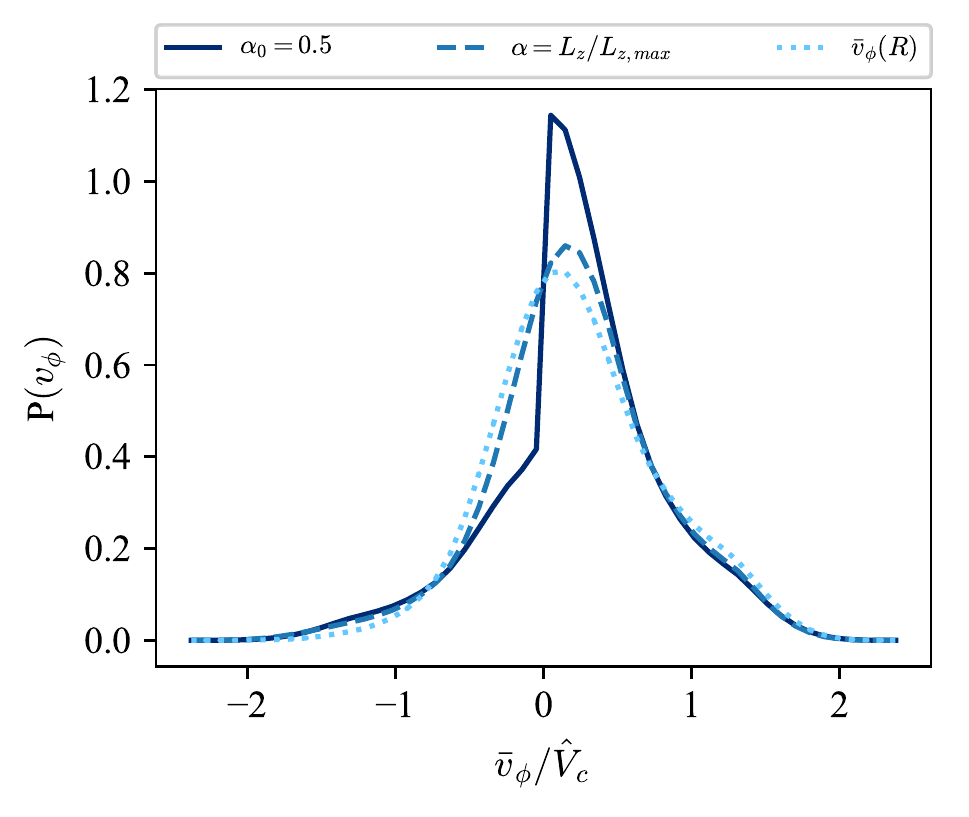}
	\includegraphics[width=0.50\textwidth]{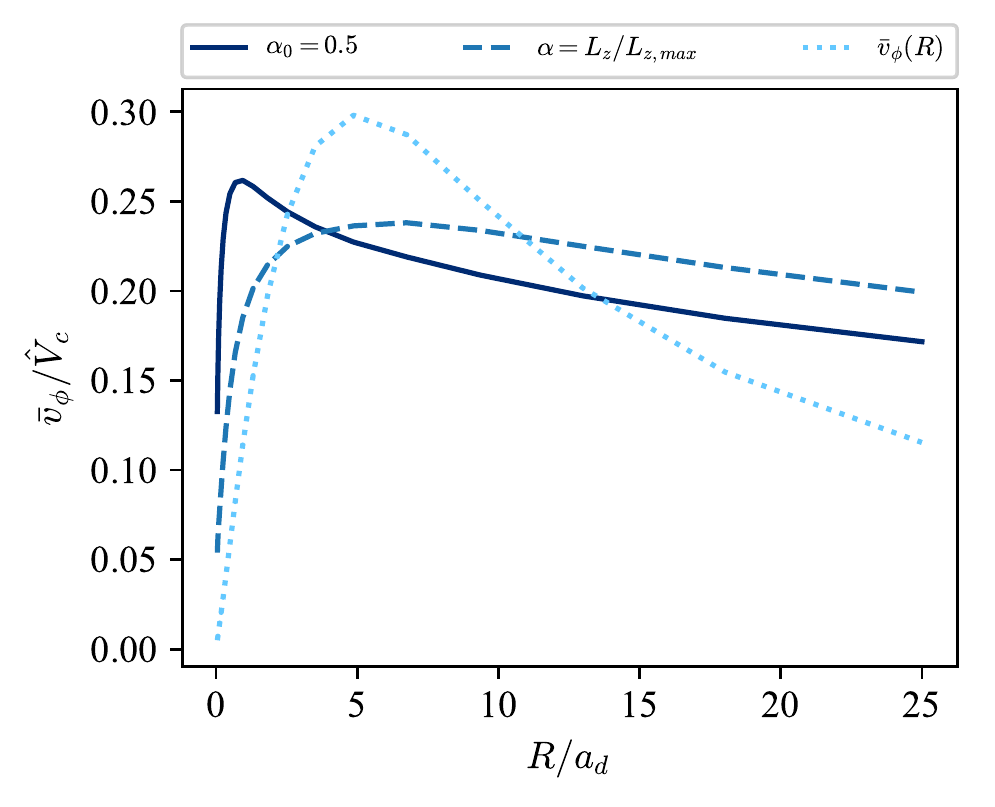}
	\caption{Azimuthal velocity probability distribution (left) and average azimuthal velocity as a function of $R$ (right) for PSDF with $f_- = \alpha f_+$ and $f_-$ computed from $\bar{v}_\phi(R)$ as defined in \eqref{eqn:rotation_profile}, using $r_a = r_s$ and $\omega$ such that that the spin parameter $\lambda(0.25r_{200}) = 0.04$.}
	\label{fig:spin_const}
\end{figure}

In order to compute the $L_z$-odd part of PSDF one needs to specify also $\bar{v}_\phi(R^2, z^2)$. While it can be, at least in principle, measured for stars or other baryonic components of galaxy, we have no way of inferring the rotational profile of the DM halo. Therefore, to address the uncertainty arising from various possible realizations of $f_-$ one can adopt the following parametrization:
\begin{align}
f_- (\mathcal{E}, L_z) = \alpha(\mathcal{E}, L_z) f_+(\mathcal{E}, L_z) \; ,
\end{align}
where $\alpha$ is an arbitrary functions that takes values in the range of $[-1,1]$ and is odd in $L_z$. One of the simplest choices is $\alpha(L_z) = \alpha_0 \cdot \textrm{sign}(L_z)$, which has been used in context of modelling stellar components of elliptical galaxies~\cite{BinneyTremaine2008}, however it introduces a discontinuity in $f(\mathcal{E}, L_z)$ at $L_z = 0$. The corresponding velocity distribution P$_{\alpha_0}(v_\phi$) is simply obtained by scaling by a constant factor the result for $\alpha_0=0$:
\begin{align}
	\textrm{P}_{\alpha_0} (v_\phi)= \textrm{P}_{\alpha_0 = 0}(v_\phi) \cdot
	\begin{cases}
		1 + \alpha_0 \; ; &  v_\phi > 0 \\
		1 - \alpha_0 \; ;& v_\phi < 0
	\end{cases} \; .
\end{align}
To avoid the discontinuity one could choose, e.g., $\alpha(L_z) = L_z / L_{z, \textrm{max}}$. While the resulting velocity distribution is smooth, it is not clear weather such $\bar{v}_\phi$ profile describes a likely configuration for DM particles or not. Therefore we also consider another option, where we assume a functional form for $\bar{v}_\phi(R,z)$. A simple choice, that was considered in the past~\cite{BinneyTremaine2008}, is the following:
\begin{align} \label{eqn:rotation_profile}
	\bar{v}_{\phi}(R) = \frac{\omega R}{1 + R^2 / r^2_a} \; .
\end{align}
It corresponds to a configuration where the system is spinning ``on cylinders", as the expression in Eq.~(\ref{eqn:rotation_profile}) is independent of $z$ (and this is convenient, since the implicit derivation of $\rho \bar{v}_\phi$ with respect to $\Psi$ in Equation \eqref{eqn:psdf_odd} does not produce additional terms). Physically, it resembles a core with solid-body rotation that diminishes towards the outskirts as $\bar{v}_\phi \propto 1/R$ for $R \gg r_a$. The comparison of $\bar{v}_\phi(R)$ and P($v_\phi$) for various rotating models is shown in Figure~\ref{fig:spin_const}. Since numerical simulations are essentially the only source of information regarding halo rotation, we follow their convention and recast halo rotation in terms of the spin parameter:
\begin{align}
	\lambda(r) = \frac{J(r)}{\sqrt{2} r M\sub{DM}(r) V_c(r)} \; ,
\end{align}
where $J(r)$ and $M\sub{DM}(r)$ are the total angular momentum and DM mass within radius $r$. We have tuned the above models to reproduce the values of spin parameter typically found in hydrodynamic simulations, $\lambda(0.25r\sub{200}) \sim 0.04$ \cite{Bryan2013}, where $r\sub{200}$ is the one of the definitions in the literature for the viral radius, and is the distance at which the average halo density within it is 200 times the critical density.

\section{Application to Milky Way}
\label{sec:MW_model}

In this section we apply the HQ method to the Milky Way and reconstruct the PSDF of its DM halo. For this purpose we use the same galactic model as in previous section, however with specific values of the parameters that were inferred for our Galaxy. The resulting PSDF, along with the assumed DM density profile, is crucial in accurately predicting the expected signals in direct DM searches (as well as in indirect searches since there are several cases in which the relevant cross-sections are velocity dependent). Furthermore, the annual modulation rate in direct detection experiments can strongly depend on the velocity anisotropy of DM particles. An even larger effect can be produced by varying the halo rotation, which also has not been previously studied in a self-consistent way at the level of PSDF.

\subsection{Mass modeling}
\label{sec:MW_parameters}

The model adopted for the Milky Way model is inspired by~\cite{Bovy2014,gala}; while the Galactic modelling has further improved in recent years, the main goal here is to highlight the trends stemming from PSDFs in an axisymmetric environment, as opposed to the spherically symmetric limit adopted so far, leaving a more detailed discussion of uncertainties and of refined mass models for future work. We therefore use the same setup introduced in Section~\ref{sec:spirals_model}, with a Miyamoto-Nagai model for the disc, a Hernquist model for the bulge and spheroidal NFW for the DM halo profile. The values of parameters used for the gravitational potentials and DM density profile are summarized in Table~\ref{tab:MW_model}. Besides these, we also fix the galactocentric distance to $R_\odot = 8$~kpc and local circular velocity $V_c(R_\odot) = 230$~km/s. The gas component and central super-massive black hole are not modeled separately, however included in our model, since we normalize the total potential according to the observed local circular velocity. To check the validity of our model we have compared it to a compilation of circular velocity data provided by \texttt{Galkin} software~\cite{Galkin} and got a match, well within the observational errors.

\begin{table}[h]
	\centering
	\begin{tabular}{l|l|l}
		Component $i$ & $\Psi_i$ ansatz & Parameters \\ \hline
		bulge & Hernquist & $x\sub{bulge} = 0.05, a_b = 1$ kpc \\
		disc & Myiamoto-Nagai & $x\sub{disc} = 0.6, a_d = 3$ kpc, $b_d = 0.28$ kpc \\
		halo & NFW & $x\sub{halo} = 0.35, r_s = 16$ kpc
	\end{tabular}
	\caption{Summary of our Milky Way model parameters. Here $x_i$ stands for the fraction of total mass in $i$-th component within the solar radius $R_\odot$, i.e. $x_i \equiv M_i(R_\odot) / M\sub{tot}(R_\odot)$ where $G M\sub{tot}(R_\odot) / R_\odot = V^2_c(R_\odot)$.}
	\label{tab:MW_model}
\end{table}

\subsection{The DM velocity distribution}

They main comparison of our results will be done against those obtained within the so-called standard halo model (SHM). As the name suggests, the latter is commonly used in approximating the phase space distribution of DM in Milky Way halo. It assumes that the PSDF can be written as the product of the DM density profile and a Maxwell-Boltzmann velocity distribution, truncated at the escape speed at a given radius in the Galaxy $v\sub{esc}(r)$:
\begin{align} \label{eqn:shm}
f\sub{SHM}(r,v) = \; & \mathcal{N} \cdot \rho\sub{NFW}(r) \cdot \exp\left( - \frac{v^2}{2 \sigma^2(r)}\right) \cdot \Theta(v\sub{esc}(r) - v) \;\;\; \\ & \textrm{where} \;\;\; \mathcal{N}^{-1} = \left(2 \pi \sigma^2 \right)^{3/2} \left( \textrm{erf} \left(\frac{v\sub{esc}}{\sqrt{2 \sigma^2}} \right) - \sqrt{\frac{2}{\pi}} \frac{v\sub{esc}}{\sigma} \exp \left( - \frac{v^2\sub{esc}}{2 \sigma^2} \right) \right) \; . \nonumber
\end{align}
The velocity dispersion $\sigma^2(r)$ can be obtained from the Jeans equation for a spherical isotropic system~\cite{BinneyTremaine2008}:
\begin{align}
\label{eqn:sigma_spherical}
\sigma{^2}(r) = \frac{1}{\rho\sub{NFW}(r)} \cdot \int_r^\infty \mathrm{d}r' \; \rho\sub{NFW}(r') \cdot \frac{\mathrm{d}\Psi}{\mathrm{d}r'}(r') \; .
\end{align}
The SHM has many shortcomings, among which the most severe one is the fact that $f\sub{SHM}(r,v)$ is built ad hoc and does not correspond to any given solution of the collisionless Boltzmann equation, except in the limit of isothermal sphere, which however does not match the DM density profiles inferred from observations. Furthermore, modern hydrodynamical simulations of structure growth indicate that DM velocity distribution significantly deviates from Gaussian and therefore models based only on the second moments of the velocity are not sufficiently accurate~\cite{Bozorgnia:2016ogo,Kelso:2016qqj,Sloane:2016kyi}.

\begin{figure}
	\includegraphics[width=0.49\textwidth]{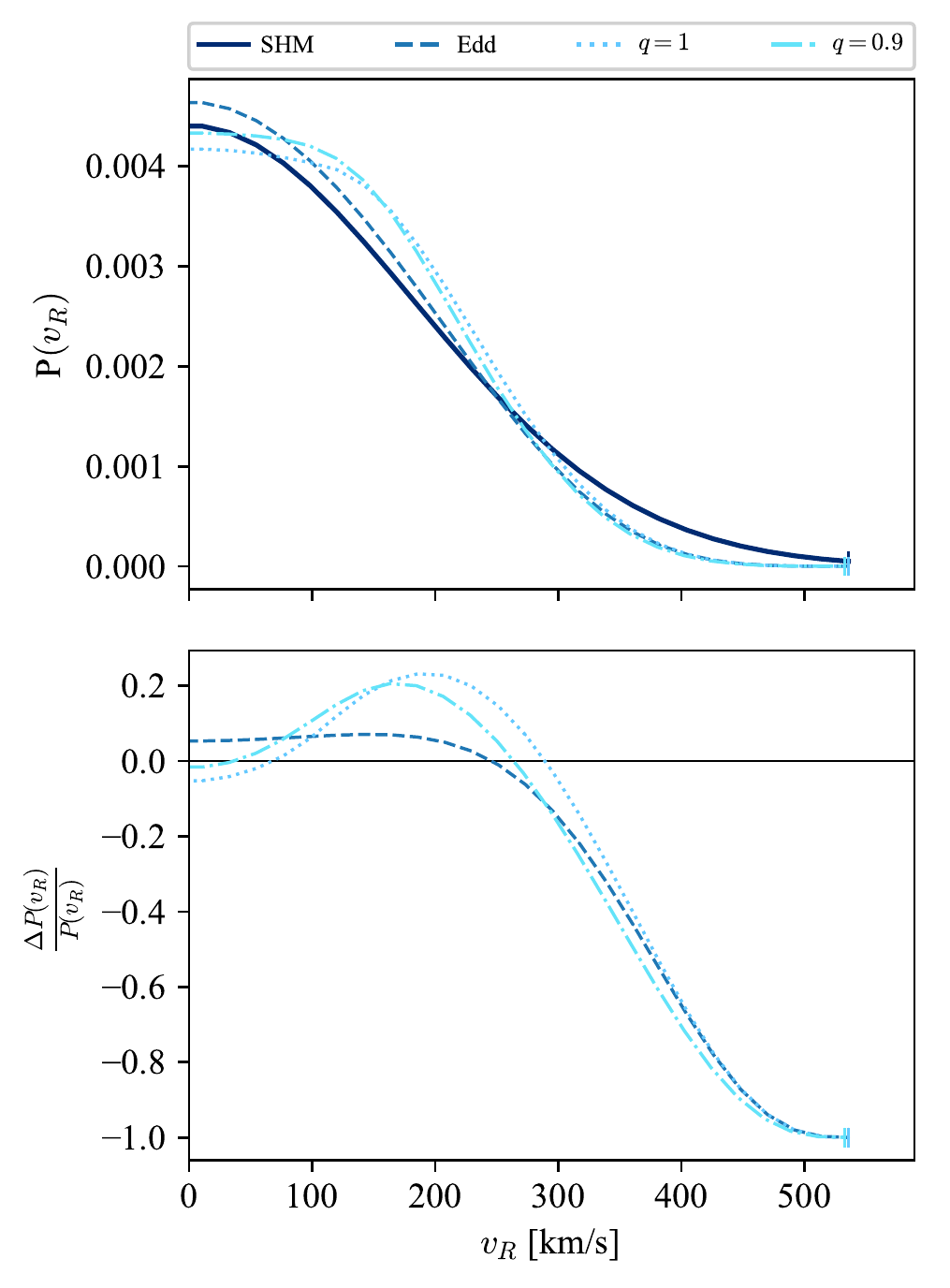}
	\includegraphics[width=0.50\textwidth]{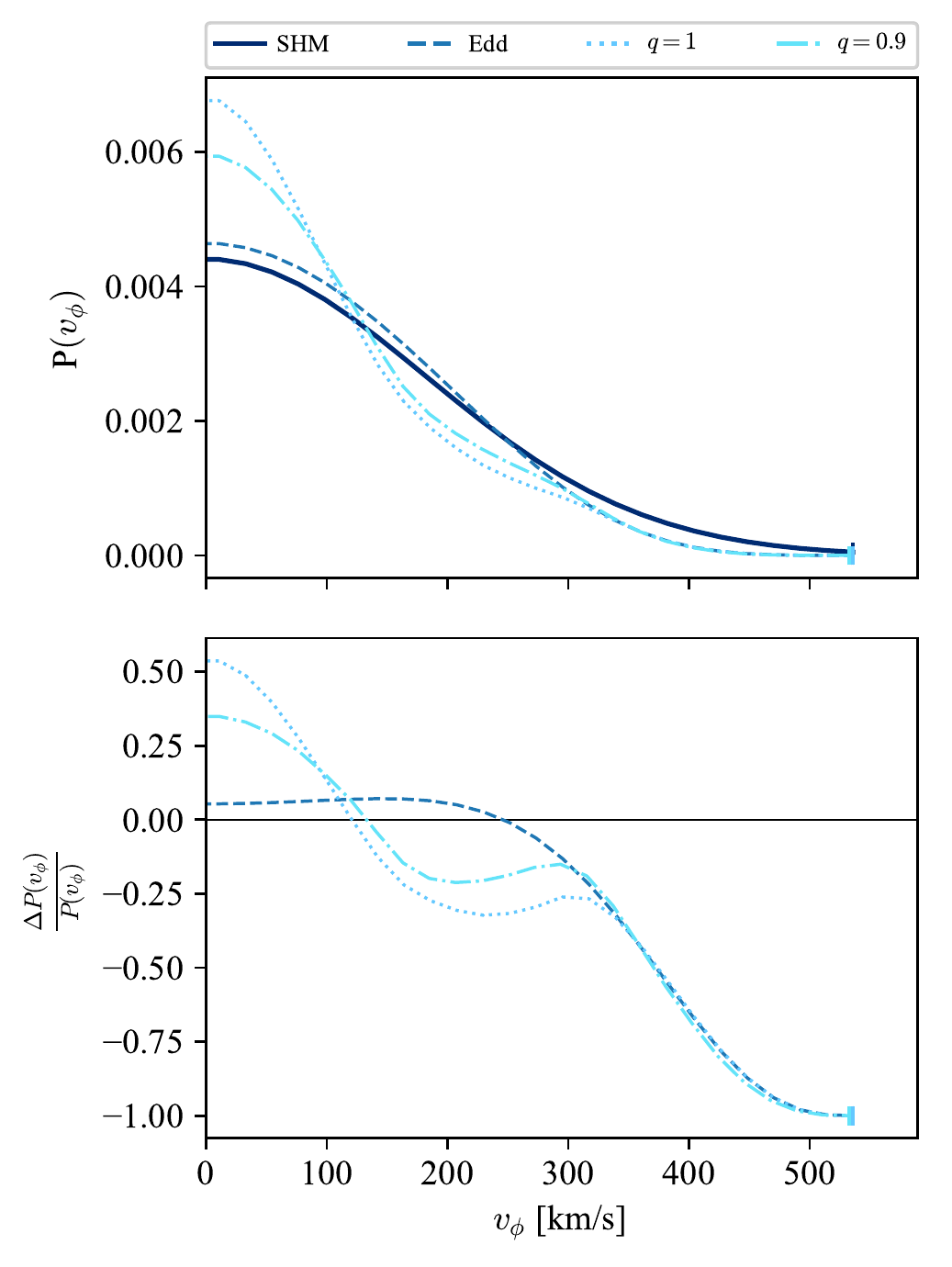}
	\caption{Local radial (left) and azimuthal (right) DM velocity distributions for SHM, spherical modelling through Eddington's inversion and axisymmetric modelling with $q=1$ and $q=0.9$, computed for the considered Milky Way sample. In the lower panels we show the relative differences with respect to the SHM.}
	\label{fig:MW_vel}
\end{figure}

\begin{figure}
	\includegraphics[width=0.50\textwidth]{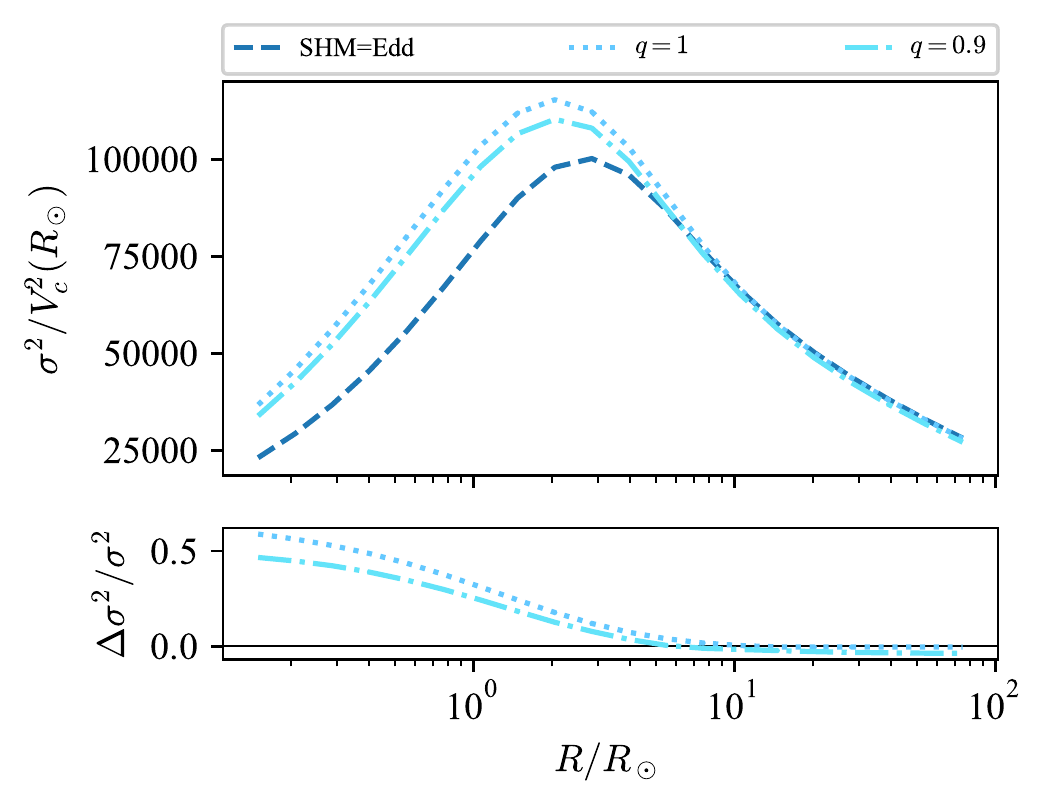}
	\includegraphics[width=0.48\textwidth]{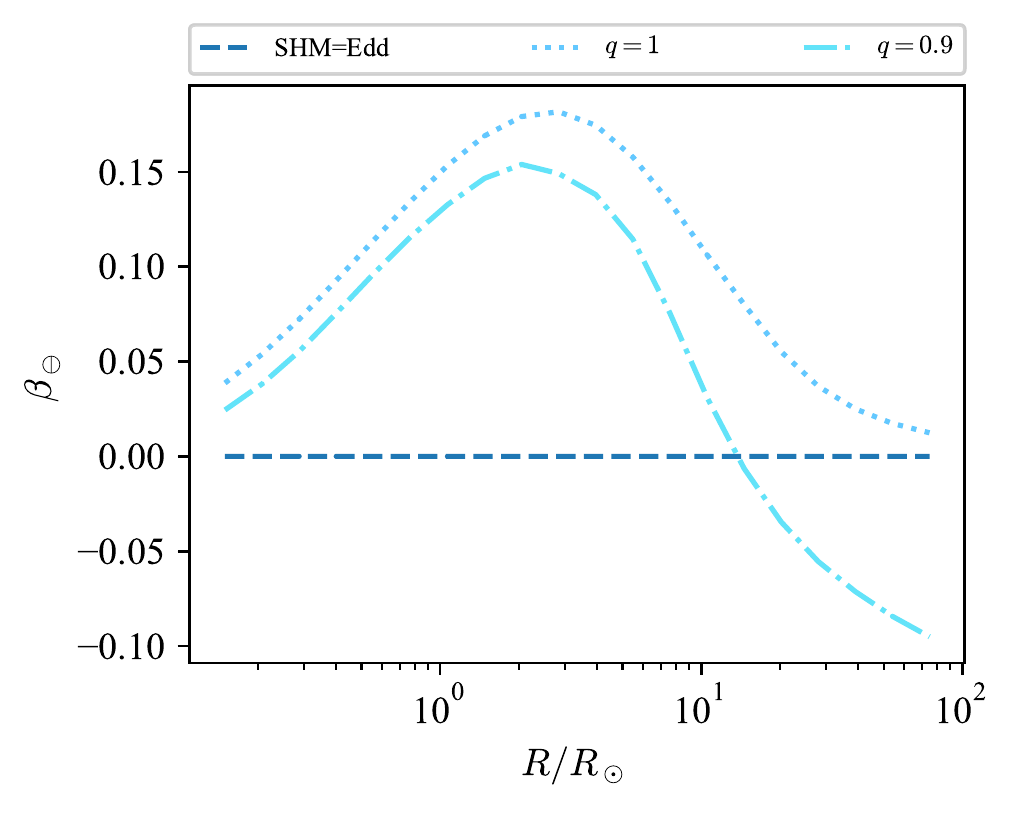}
	\caption{Local DM velocity dispersion (left) and anisotropy (right) as function of galactocentric distance for SHM (the results for spherical modelling through Eddington's inversion are equivalent to the SHM ones, since we compute the velocity dispersion for the latter according to the Equation~\eqref{eqn:sigma_spherical}) and axisymmetric modelling with $q=1$ and $q=0.9$, computed for the considered Milky Way sample. In the lower left panel we show the relative difference of velocity dispersion with respect to the SHM.}
	\label{fig:MW_disp}
\end{figure}

To properly address the Milky Way-like structure, presented in Section~\ref{sec:MW_parameters}, one needs to resort to the HQ axisymmetric generalization of Eddington's approach. In Figure~\ref{fig:MW_vel} we highlight the main differences in the local DM velocity distributions that arise when using different PSDF models, namely the SHM $f\sub{SHM}(R_\odot,v)$, the spherical limit obtained by implementing Eddington's formula $f\sub{Edd}(\mathcal{E})$, and the HQ model $f\sub{HQ}(\mathcal{E}, L_z)$ with $q=1$ and $q=0.9$. As already discussed in previous section, the radial velocity distribution of DM is shifted towards higher velocities when properly accounting for an axisymmetric disc, however such effect can be compensated by flattening the halo. For the azimuthal velocity distribution, trends are the opposite; the axisymmetric disc modelling leads to a velocity distribution that is skewed towards lower velocities, while flattening the DM halo shifts it towards higher velocities. Since there is evidence for only a mild flattening of the Milky Way halo, the effect of the axisymmetric disc prevails, even when using the lower bound of inferred values of $q$. Moreover, the resulting velocity distributions significantly deviate from Gaussian shapes, despite the aforementioned apparent compensation between the effects of stellar disc and halo flattening, in further support to the need of going beyond the SHM. The axisymmetric modelling also introduces changes in the velocity dispersion and velocity anisotropy in the galactic plane, as displayed in Figure~\ref{fig:MW_disp}. For axisymmetric models, the total velocity dispersion, shown in the left hand side plot, increases in the central part of the Galaxy, while in the outskirts it reduces to the case of spherical halos, as the effect of the disc becomes negligible, while it is slightly lower for oblate halo as in this case there is a lower total mass. In agreement with the trends seen in velocity distribution functions, when DM halo is assumed to be spherical, one gets a radially biased anisotropy at all radii, while for the flattened model there is a radial biased for $R \lesssim R_\odot$ and an azimuthal bias for $R \gtrsim R_\odot$, as shown in the plot on right hand side of Figure~\ref{fig:MW_disp}. 

\section{Impact on Direct Detection searches}
\label{sec:DD}

The local distribution of DM is one of the crucial unknowns in constraining the DM interaction rate with baryonic matter through direct detection experiments. The latter are designed to measure nuclear recoils due to scattering of halo DM particles on the nuclei in target materials. The signal is typically quantified as differential event rate:
\begin{align}
\frac{\mathrm{d}R}{\mathrm{d}E_r} = \frac{1}{m_A m_\chi} \cdot \int_{|\vec{v}| > v_\textrm{min}} \mathrm{d}^3v \; f(\mathcal{E}, L_z) \cdot v \cdot \frac{\mathrm{d}\sigma}{\mathrm{d} E_r} \label{eqn:rate_dif} \\
v_{\textrm{min}} = \sqrt{\frac{m_A E_r}{2 \mu^2_{A \chi}}} \; \; \; , \; \; \; \mu_{A \chi} = \frac{m_A m_\chi}{m_A + m_\chi} \nonumber
\end{align}
where $R$ is the event rate, $E_r$ the recoil energy, $m_{A/\chi}$ the nucleus/DM mass, $v = |\vec{v}|$ the velocity of DM particle in the detector (LAB) frame and $\frac{\mathrm{d}\sigma}{\mathrm{d} E_r}$ the corresponding differential cross-section. For spin-independent (SI) interactions the latter can be written as:
\begin{align}
\frac{\mathrm{d}\sigma}{\mathrm{d} E_r} = \frac{m_A \sigma^{\textrm{SI}}_{n}}{2 \mu^2_{A \chi} v^2} A^2 F^2(E_r) \; ,
\end{align}
where $\sigma^{\textrm{SI}}_{n}$ is the SI cross-section at zero momentum transfer, $A$ the mass number of target nucleus and $F(E_r)$ the corresponding (energy dependent) form factor. As can been seen from the above expression, SI differential cross-section introduces an additional factor of $v^{-2}$, which also appears in the case of spin-dependent (SD) interactions, however this is not always true for more general scattering operators. For SI and SD differential cross-sections one can factorize Eqn.~\eqref{eqn:rate_dif} into a constant term stemming from the specific particle physics model times an integral that is determined by the DM velocity distribution:
\begin{align}\label{eqn:g}
g(v_\textrm{min}) \equiv \frac{1}{\rho_\odot} \int_{|\vec{v}| > v\sub{min}} \mathrm{d}^3v \; \frac{f(\mathcal{E}, L_z)}{v} \; .
\end{align}
where $\rho_\odot$ is the local DM density. When considering a broader range of effective two-to-two scattering operators one generically encounters cross-section terms with an additional power of $v^2$ and therefore it is useful to define also:
\begin{align}\label{eqn:h}
h(v_\textrm{min}) \equiv \frac{1}{\rho_\odot} \int_{|\vec{v}| > v\sub{min}} \mathrm{d}^3v \; f(\mathcal{E}, L_z) \cdot v \; .
\end{align}
In the above integrals one must note that Earth, together with DM detectors, is moving with respect to the Galactic rest frame with velocity $\vec{v}_\oplus(t)$, which is a sum of the local circular velocity $V_c(R_\odot)$, the peculiar motion of the Sun~\footnote[1]{In this work we adopt $U_\odot = 11$, $V_\odot = 12$ and $W_\odot = 7$ km/s, consistent with~\cite{Schonrich2009}, where $\hat{U}$ points towards the galactic center, $\hat{V}$ in positive direction of the galactic rotation and $\hat{W}$ towards the galactic north pole.} and the Earth velocity relative to the Sun~\footnote[2]{For Earth velocity relative to the Sun we use $|\vec{v}\sub{E}| = 30$ km/s, with the orbit tilted by $60°$ about the radial axis with respect to the galactic plane. For a more detail treatment see, e.g., \cite{Lee2013,McCabe2014}.}. Therefore, the relative energy and angular momentum entering the PSDF become:
\begin{align}
	\mathcal{E} = \Psi(R^2_\odot, 0) - (\vec{v} + \vec{v}_\oplus^2) / 2 \;\;\; \textrm{and} \;\;\; L_z = R_\odot \left( (\vec{v} + \vec{v}_\oplus) \cdot \hat{e}_\phi \right) \; ,
\end{align}
where $\hat{e}_\phi$ is the unit vector pointing along the azimuthal direction. Since the largest contribution to $\vec{v}_\oplus(t)$ comes from the co-rotation of Local Standard or Rest (LSR) with galactic disc, this introduces a strong anisotropy along the azimuthal direction, which can conspire with the axisymmetric PSDF modeling to produce significant deviations with respect to the isotropic approach.

\begin{figure}
	\centering
	\includegraphics[width=0.49\textwidth]{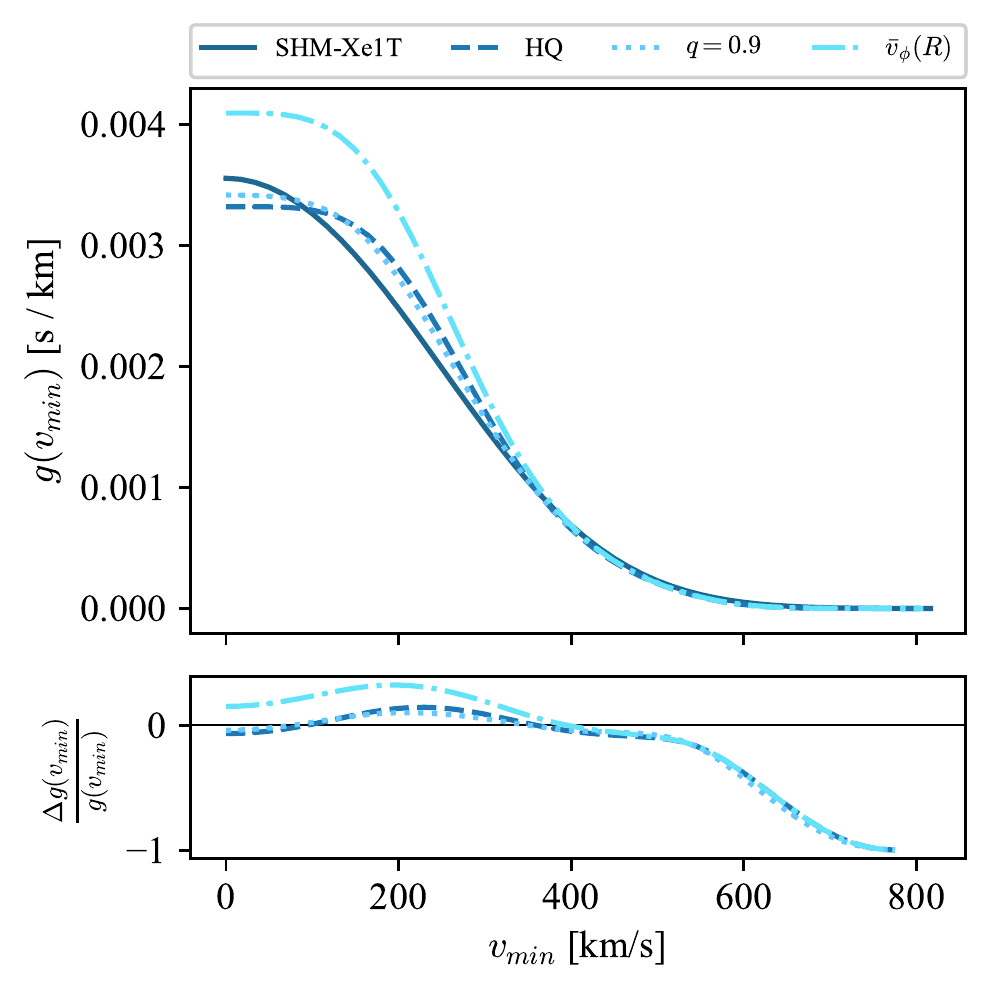}
	\includegraphics[width=0.49\textwidth]{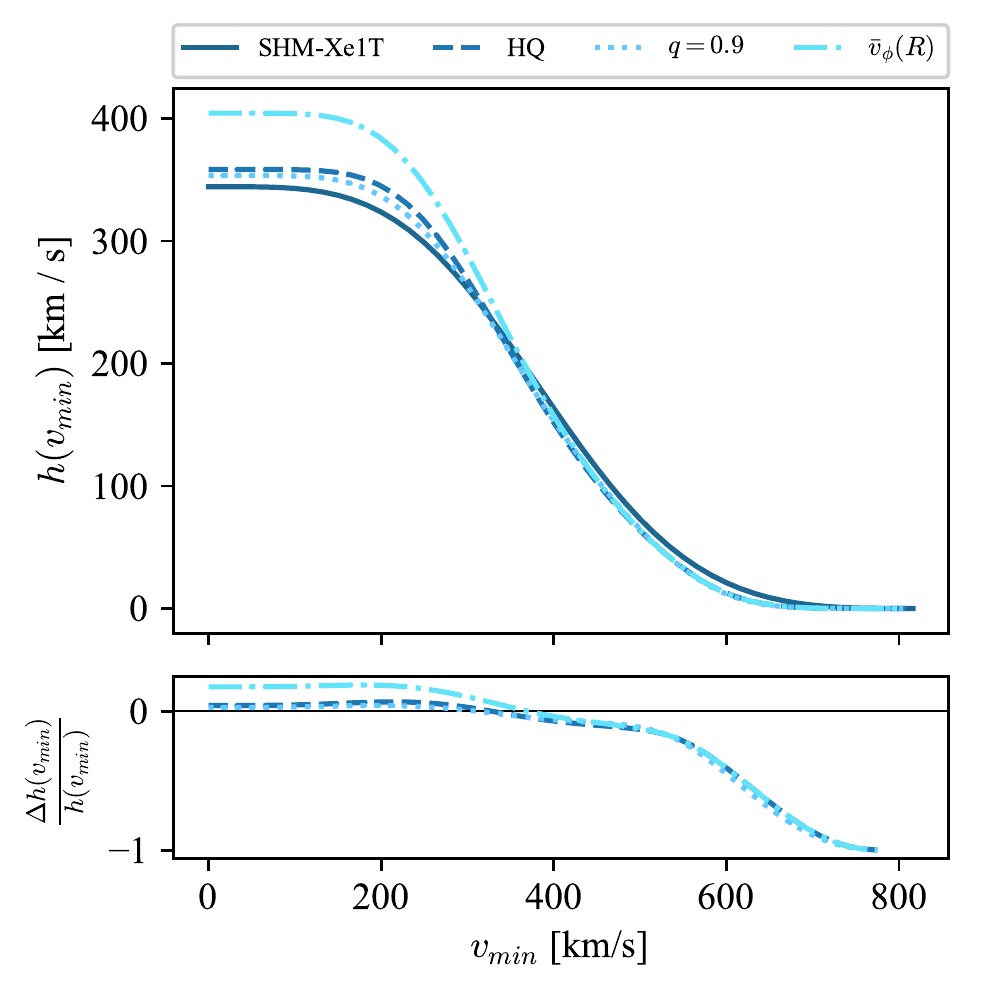}
	\caption{Astrophysical factors defined in Equations~\eqref{eqn:g} and~\eqref{eqn:h}, which appear in the deferential event rate, as a function of minimal scattering velocity for SHM and axisymmetric model with $q=1$ (HQ), $q=0.9$  and spherical rotating halo with $\bar{v}_\phi(R)$ as defined in Equation \eqref{eqn:rotation_profile}, using $r_a = r_s$ and $\omega$ such that the spin parameter $\lambda(0.25r_{200}) = 0.04$.}
	\label{fig:eta}
\end{figure}

In Figure~\ref{fig:eta} we show $g(v\sub{min})$ and $h(v\sub{min})$ for the SHM adopted by Xenon1T collaboration to present their results~\cite{Xenon1T:2018} (SHM-Xe1T) and a set of axisymmetric models that differ in flattening and rotation. The first notable difference is the fact that the SHM-Xe1T has a somewhat higher local escape velocity, $v\sub{esc} = 544$ km/s, compared to our axisymmetric models, which have $v\sub{esc} \approx 535$ km/s, as well as a lower local DM density, $\rho_\odot = 0.3$ GeV/cm$^3$, with respect to $\rho_\odot \approx 0.35$ GeV/cm$^3$ that we find for the sample Milky Way model described in Section~\ref{sec:MW_parameters} (we give approximate values since both quantities slightly depend on $q$). Furthermore, the SHM is sharply truncated at $v\sub{esc}$, while the velocity distributions in other models we display fall smoothly to zero, leading to larger $g(v\sub{min})$ and $h(v\sub{min})$ for SHM-Xe1T at large values of $v\sub{min}$. For smaller values of $v\sub{min}$ the axisymmetric models take over and the rotating halo gives the largest values for $g$ and $h$. This follows from the fact that the total velocity dispersion increases for the axisymmetric models and therefore they increase the number of ``high velocity scatterings", which contribute the most to integrals in Eqns.~\eqref{eqn:g} and \eqref{eqn:h}. For co-rotating halo, corresponding to the azimuthal velocity profile defined in Equation~\eqref{eqn:rotation_profile} with $\omega$ that yields $\lambda(0.25r_s)=0.04$ and assuming $r_a = r_s$, the functions take even larger values as the the convolution in aforementioned integrals is performed closer to the peak of $f(\mathcal{E}, L_z)$, while the effect for counter-rotating halo would be the opposite.
 
\begin{figure}
	\centering
	\includegraphics[width=0.49\textwidth]{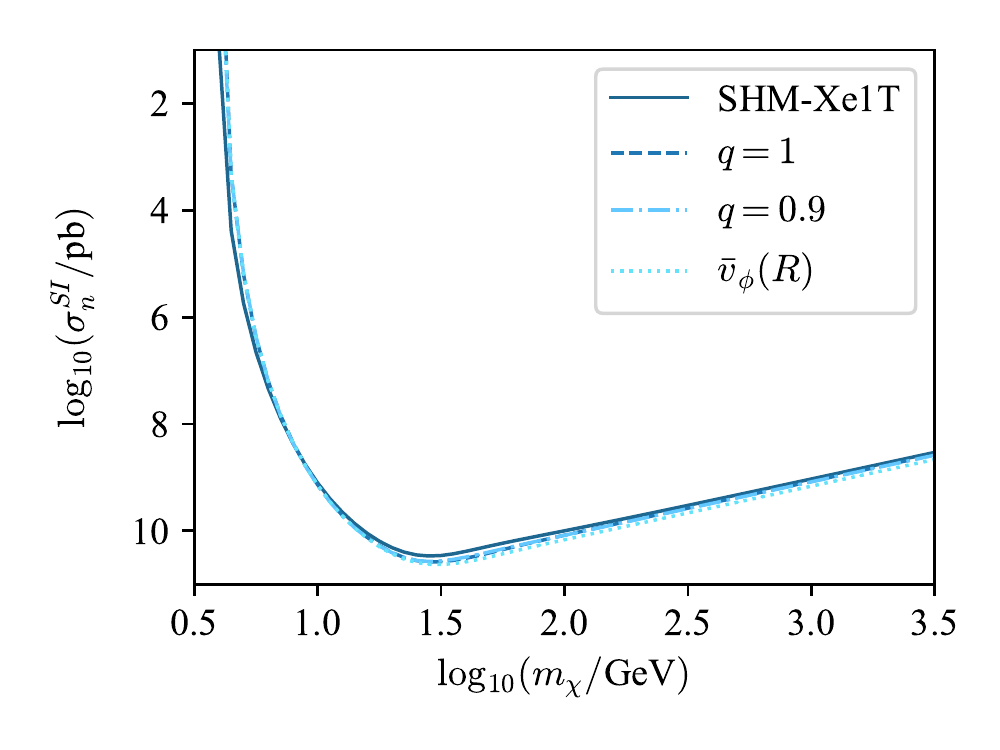}
	\includegraphics[width=0.49\textwidth]{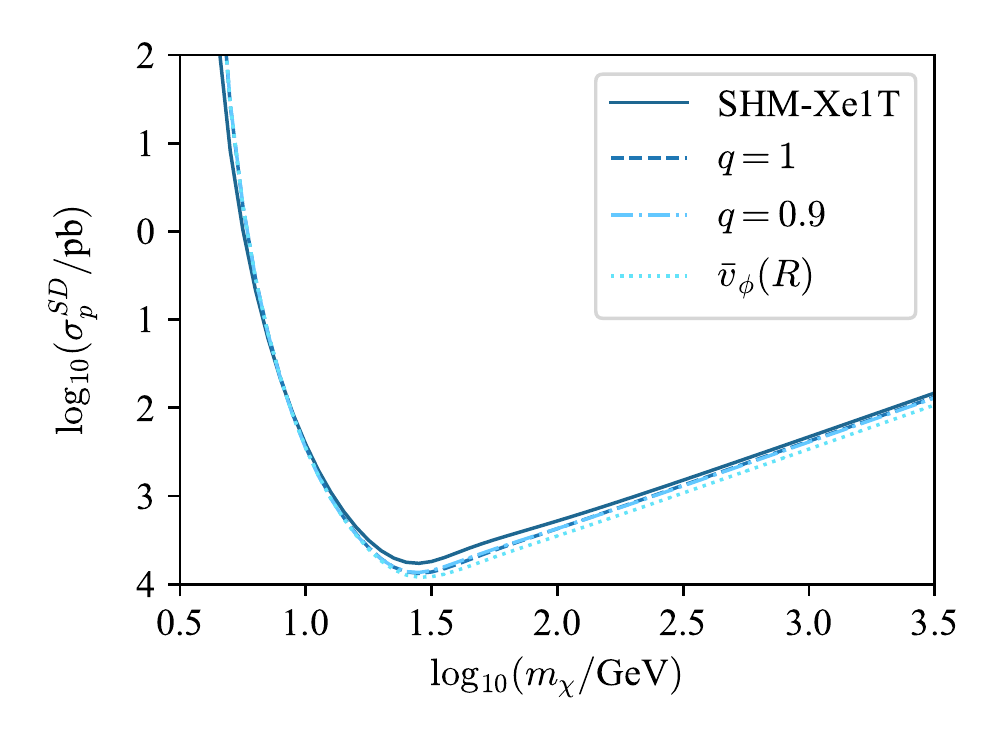}
	\caption{Xenon1T cross-section -- mass exclusion plots for SHM and axisymmetric modelling with $q=1$, $q=0.9$ and spherical rotating halo with $\bar{v}_\phi(R)$ as defined in Equation \eqref{eqn:rotation_profile}, using $r_a = r_s$ and $\omega$ such that the spin parameter $\lambda(0.25r_{200}) = 0.04$. Left plot shows the results for spin-independent and right plot for spin-dependent scattering.}
	\label{fig:sigma_m}
\end{figure}

In Figure~\ref{fig:sigma_m} we show the resulting exclusion plots of DM-nucleon cross-section as a function of DM mass for the standard spin-independent (SI) and spin-dependent (SD) scattering computed from the Xenon1T null results~\cite{Xenon1T:2018} using the \texttt{DDCalc} software \cite{DDCalc}. As could be anticipated from the trend seen in $g(v\sub{min})$, rotating halos give the strongest limits, except at DM masses below $\sim 20$ GeV where the SHM-Xe1T gives an overestimated constraints due to the (artificial) pile up of high velocity particles close to the escape velocity. Non-rotating axisymmetric models yield predictions much closer to the SHM-Xe1T for $m_\chi \gtrsim 20$ GeV and therefore the exclusion limits remain almost unchanged when viewed on the logarithmic scale. There are however some minor differences around the kink at $m_\chi \sim 30$ GeV, where the sensitivity of Xenon experiments is the highest, further strengthening the case for proper axisymmetric modelling. By examining $h(v\sub{min})$, which is suppressed for the SHM-Xe1T with respect to all the other considered models for $v\sub{min} \lesssim 400$ km/s, we expect more significant changes in the case of other possible DM-nucleus scattering operators, which include additional powers of velocity dependence. We plan a detail investigation of exclusion limits for a general set of elastic, as well inelastic, scattering cross-sections in a subsequent work.

\begin{figure}
	\centering
	\includegraphics[width=0.50\textwidth]{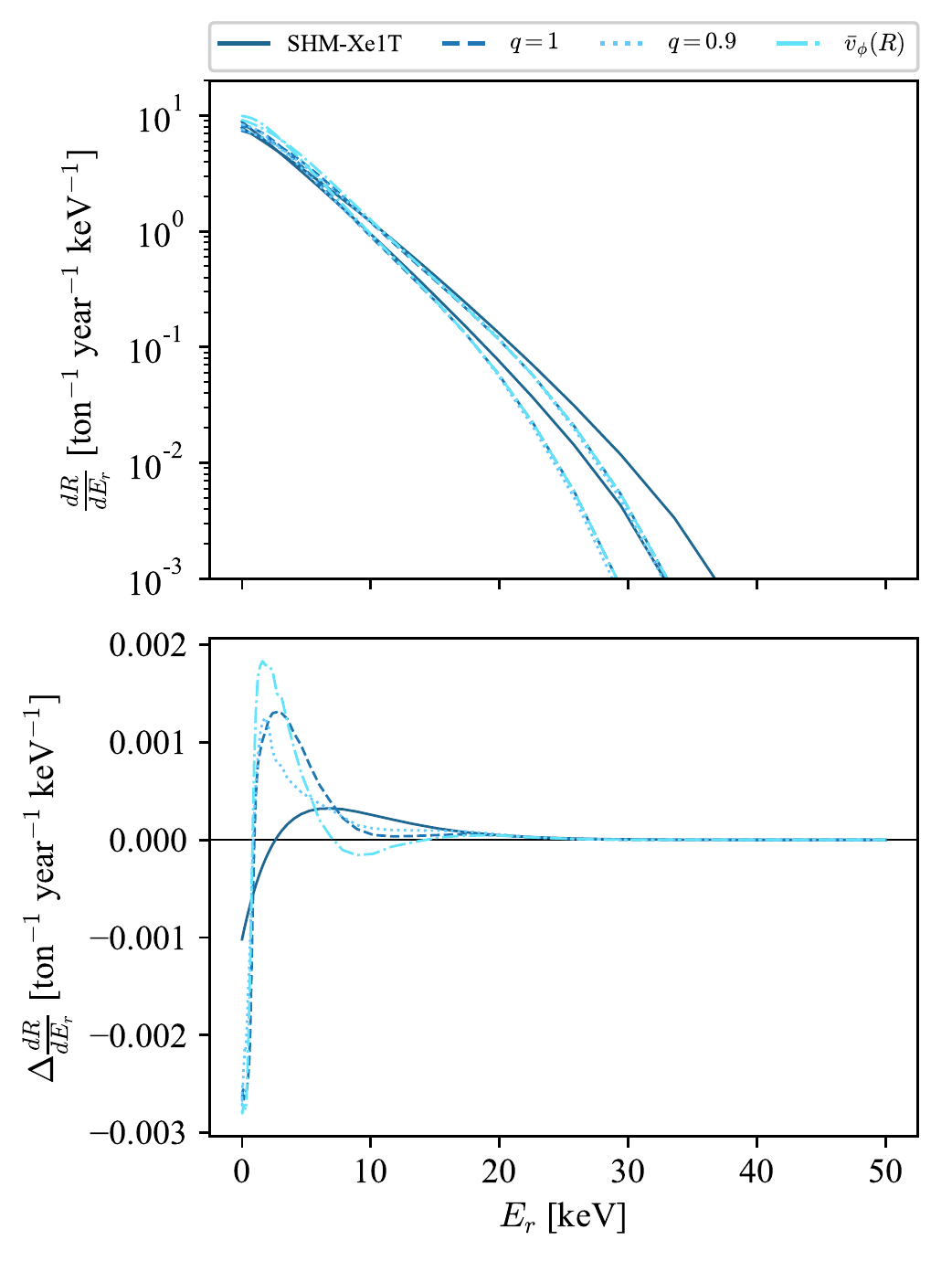}
	\includegraphics[width=0.49\textwidth]{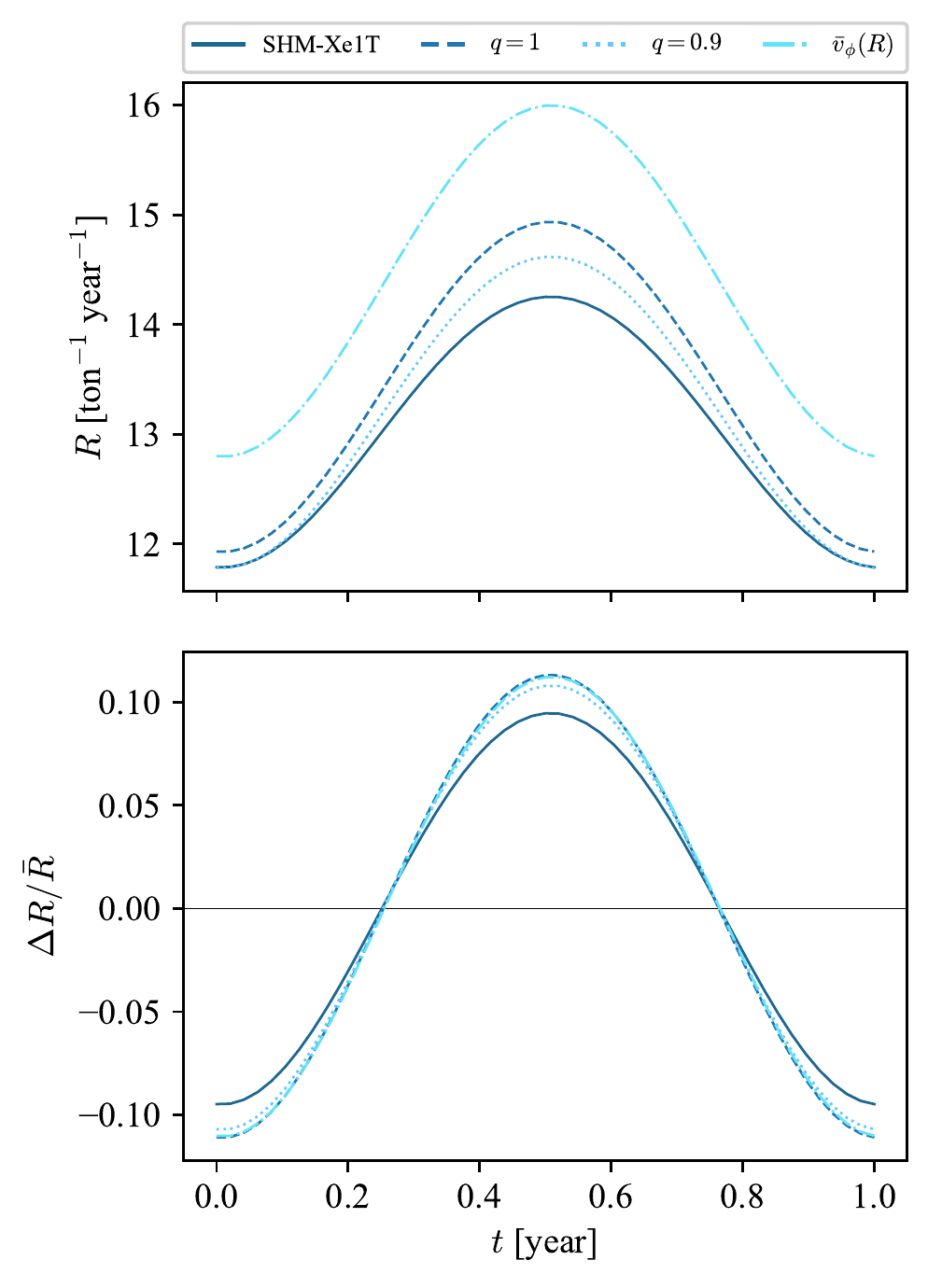}
	\caption{\textit{Left:} Annual modulation of the differential scattering rate as a function of recoil energy. Top panel shows the two extreme cases with respect to the time of the year, while the lower plot shows the absolute difference between the two. \textit{Right:} Annual modulation of the event rate as a function of time. Upper panel shows the absolute value, while the lower panel shows the relative value with respect to the annual average for the given model. The results are computed for $^{131}$Xe, assuming SI corss-section $\sigma_n^\textrm{SI}=10^{-46}\textrm{cm}^2$ and $m_\chi=20$ GeV, while the considered PSDF models are the same as in Figure~\ref{fig:sigma_m}.}
	\label{fig:modulation}
\end{figure}

Finally, we apply the self-consistent axisymmetric modelling of galactic DM phase-space distribution to predict the annual modulation in DM-nucleon scattering rates. This is particularly interesting in the light of well established DAMA/LIBRA anomaly~\cite{Bernabei2016,Bernabei2018}, which is, however, very difficult to reconcile with null results of other direct detection experiments. In Figure~\ref{fig:modulation} we present our findings regarding the impact of axisymmetric halo modelling on the annual modulation rate. In the plots on left hand side we show the yearly maximal and minimal differential scattering rates (above) and their difference (below) for the set of considered models. As expected, with find again the largest difference for co-rotating halos, followed by the $q=1$ model. The aforementioned cancellation of effects between axisymmetric modelling and flattening of the halo drives the $q=0.9$ case closer to the SHM-Xe1T, which predicts the smallest modulation signal. On the right side of Figure~\ref{fig:modulation} we show the absolute scattering rate for energy bin $E_r \in [5\textrm{keV}, 15 \textrm{keV}]$ as a function of time, assuming $100\%$ detection efficiency in this range. Besides the over-all shift in the rates, which is essentially analogous to the change in limits on SI and SD cross-sections, there is also an appreciable difference in the modulation amplitude. It is the most pronounced for axisymmetric modelling with spherical halos, leading to roughly $25$\% larger modulation amplitude then the SHM, while it is somewhat smaller for flattened halo. We note that this result depends on the energy binning, as can be clearly seen from the difference in differential event rates in the lower plot on the left hand side of Figure~\ref{fig:modulation}. 

\section{Conclusion}
\label{sec:conclusions}

In this paper we have addressed the problem of reconstructing the phase-space distribution function for an extended collisionless system, whose density profile is known and which is in equilibrium within an axisymmetric gravitational potential, with a large contribution from a strongly flattened component. This picture is relevant in several contexts; we have applied it to derive the phase-space distribution function for dark matter particles in the halo of spiral galaxies, with particular attention to the Milky Way and its implications for dark matter direct detection. 

The method we have implemented relies on the assumption that the distribution function depends only on energy and on $L_z$, the component of the angular momentum parallel to the axis of symmetry. Within such a model, for a given gravitational potential, the relation which gives the density as an integral over velocities of the distribution function can be uniquely inverted. This allows to self-consistently retrieve the dark matter phase-space distribution function for any axisymmetric galactic mass model decomposition, as derived from dynamical observations. The procedure was originally proposed by Lynden-Bell and later refined in its numerical implementation by Hunter \& Qian; it has been applied here for the first time to cases with a thin stellar disc, after overcoming a technical difficulty which has probably prevented its use before. 

Our approach is a generalization of the Eddington's inversion formula, which is valid for spherically symmetric systems and is much less demanding from a computational point of view. While Eddington's formula has been applied to the Milky Way quite extensively before, we have shown here that the spherically symmetric approximation fails to address important features that are due to the Milky Way axisymmetic structure, in particular at the location of the Sun and in the inner Galaxy. The main feature is that the pressure felt by the collisionless system in the axisymmetic potential is different in the meridional plane with respect to the azimuthal direction. Compared to the other phenomenological models, we have also noticed a strong deviation from the case when the velocity distribution is described by a simple truncated Gaussian.

In Section \ref{sec:axi_modelling} we have introduced the method and discussed some general results concerning the distribution function for DM particles in spiral galaxies. We have found that velocity distributions differ significantly when the shape of stellar component is changed from a spherical profile to a disc, despite keeping the total mass profile of the galaxy unchanged. The radial velocity distribution tends to be shifted towards higher velocities, while the azimuthal component gains power at low velocities. We obtain similar results when varying the fraction of total mass contained in the stellar component while keeping constant circular velocity at a chosen characteristic radius. Besides highlighting the need for axisymmetric modelling, this results also show that it is crucial to correctly asses the mass model decomposition, rather then just the overall normalization of the potential well. The latter, however, is also important, since it determines the escape velocity of the system and plays a crucial role in correctly estimating limits on the scattering cross-section for light DM candidates. Besides the axisymmetric modelling of the stellar component, we have also considered oblate and prolate DM halos, with again a non-negligible impact on the velocity distribution. Flattening the DM profile leads to the opposite effect with respect to making the axisymmetric stellar component heavier and can, to some degree, compensate the changes in the velocity distributions. The opposite is found in case of prolate halos, where the radial velocity distribution getting shifted to higher velocities while the azimuthal component becoming even more peaked at low velocities. Finally, we have considered rotating dark halos, exploiting the fact that for any give rotational velocity profile one can uniquely determine the $L_z$-odd part of the phase-space distribution function. We have adopted various possible ways of modelling the halo rotation, either by simply constructing the $L_z$-odd part of the PSDF from $L_z$-even part, or computing it for an assumed parametric rotation velocity profile.

In second part of our work we have focused on the study of Milky Way and re-analyzing the direct detection predictions within an axisymmetric setting. We have used a sample Milky Way model, composed of stellar bulge, disc and dark matter halo, which reproduces the galactic rotation curve. This picture has been sufficient to highlight the main effect of the presence of a thin stellar disc, compared to the usually adopted spherical approximations, namely the so-called standard halo model and the distribution function obtained through Eddington's inversion formula. As benchmark distribution functions we have implemented two realizations of axisymmetric models, with a spherical and a slightly flattened dark matter halo, illustrating the relevant differences compared to the fully spherical models. 

Finally, we have discussed the impact on direct detection signals, first at the level of generic interactions, encoding the dependence of the signal on dark matter velocity distributions, and then going to an example case for detection rates and annual modulation effects. The corresponding bounds on spin-independent and spin-dependent cross-sections are roughly 50\% stronger for rotating halo, while the effects are smaller for other models, except at low $m_\chi$ where the standard halo model over-predicts the rate of scattering due to relatively large abundance of high velocity particles that is induced by a sharp truncation of the velocity distribution. For what regards the annual modulation signal, there is both an enhancement in the expected event rate and a sizable change in the modulation amplitude, which can be roughly 25\% larger with respect to the standard halo model.

In our future work we plan to use the introduced formalism to further explore mass modelling of the Milk Way and perform a more comprehensive study of the impact on direct detection, including additional velocity dependent operators, as well as the case for inelastic scatterings. Results of this paper are also readily applicable to indirect detection of DM particles with velocity dependent pair-annihilation rates. Moreover our approach is of interest also in wider context, such as for studying the dynamics of halo stars in spiral galaxies or for refining estimates of gravitational lensing searches for primordial black holes.

\section*{Acknowledgments}

We thank P. Salucci for useful discussions. Our work was partial support from the European Union's Horizon 2020 research and innovation programme under the Marie Sk\l odowska-Curie grant agreement No 690575 and from the European Union's Horizon 2020 research and innovation program under the Marie Sk\l odowska-Curie grant agreement No 674896.

\appendix

\section{Difficulties on the PSDF evaluation with the HQ method} \label{app:branch_cuts}

The HQ method is rather efficient in numerically computing phase-space distribution functions for isolated self-gravitating populations, i.e. when the gravitational potential $\Psi(R^2,z^2)$ is self-consistently generated by the system density profile $\rho(R^2,z^2)$. In this case, the proposal by Hunter \& Qian, as reproduced in Eqns.~\eqref{eqn:contour_finite} and \eqref{eqn:contour_infinite}, for the contour $C(\mathcal{E})$, entering critically in the evaluation of the integral in Eqn.~\eqref{eqn:psdf}, is a good choice. The reason for which it works well is that it generally avoids the inclusion of additional singularities and/or crossing of branch cuts which the analytic continuation of $\frac{\mathrm{d}\rho}{\mathrm{d}\Psi}$ in the complex plane may introduce (see the discussion in \cite{HQ1993} for details). However, when addressing configurations in which the total gravitational potential is not entirely sourced by the density under consideration, but there are also additional contributions from other components, the method may encounter difficulties. For example, in the case considered in this paper of a DM halo combined with an external Myamoto-Nagai potential, Eqn~\eqref{eqn:potential_MN}, an additional branch cut in $\Psi(R^2,z^2)$ occurs along the real axis at $z^2 < - b_d^2$. As a consequence the Jacobian for the change of variables from $\rho(R^2, z^2)$ to $\rho(R^2, \Psi)$ contains a discontinuity, due to which it is not always possible to invert the potential at every point $\xi(\theta)$ along the contour in Eqns.~\eqref{eqn:contour_finite} or \eqref{eqn:contour_infinite}. The inverse for $\xi(\theta=0)$ does exist by construction, however for larger values of $\theta$, there is no guarantee that one can find $z^2$ such that:
\begin{align}
\xi(\theta) = \Psi \left(\frac{L_z^2}{2(\xi(\theta) - \mathcal{E})}, z^2 \right) \;.
\end{align}
This indeed does not happen for certain values of $\mathcal{E}$ and $L_z$, depending also on the choice of the ``thickness" of the contour $h$. It is sometimes possible to mitigate the problem by adjusting the value of $h$, or by choosing a different contour shape, in order to avoid values of $\xi$ for which the inversion breaks down. For finite potentials an alternative choice of the upper half of the contour is a boxy path parametrized by:
\begin{align} \label{eqn:contour_box}
\xi_1(s) & = \psi_{\textrm{env}} + i \, h \, s \; ,\\
\xi_2(s) & = \psi_{\textrm{env}} \, (1 - s) + i \, h \; , \\
\xi_3(s) & = i \, h \, (1 - s) \; ,
\end{align}
with $s \in \left[0, 1\right]$. Again, by tuning the parameter $h$ one can try to avoid the values of $\xi$ where the inversion fails. If this is not possible even along this second path (e.g. it requires again such a small $h$ that one faces loss of numerical precision when integrating around the pole), an approximate solution is to pick out $h$ in such a way that the discontinuity occurs along $\xi_3(s)$. Indeed, it turns out that the contribution to the integral along $\xi_3(s)$ is negligible, being orders of magnitude smaller then the ones from $\xi_1(s)$ and $\xi_2(s)$, as it approaches the value of potential at infinity (i.e. $\Re \left[\xi_3(s) \right] = 0$, $\forall s$), where $\rho(R^2, z^2)$ and it's derivatives vanish. We explicitly checked that $f(\mathcal{E}, L_z)$ computed through this approximation successfully reproduces the initial density. Furthermore, in practice the error of neglecting the $\xi_3(s)$ contribution is much smaller then the errors coming from the numerical integration along the rest of the contour.

\bibliographystyle{apsrev}
\bibliography{library.bib}
	
\end{document}